%% file: paper.tex
\documentclass[
  aps,prb,
  twocolumn,
  superscriptaddress,
  floatfix,
  longbibliography,
  amsmath,amssymb
]{revtex4-2}

\usepackage{microtype}
\input{macros}

\begin{document}
\input{body}

\bibliography{My_Library,refs}
\end{document}

%% file: macros.tex
\usepackage{graphicx}
\usepackage{bm}
\usepackage{mathtools}   
\usepackage{xcolor}
\usepackage[caption=false]{subfig}
\usepackage[colorlinks,linkcolor={red!50!black},citecolor={blue!50!black},
            urlcolor={blue!50!black}]{hyperref}

\graphicspath{{figures/}}

\newif\ifdraftmode\draftmodetrue
\ifdraftmode
  \newcommand{\todo}[1]{\textcolor{red}{\textsf{\footnotesize [TODO: #1]}}}
  \newcommand{\note}[1]{\textcolor{blue}{\textsf{\footnotesize [#1]}}}
  \newcommand{\caveat}[1]{\textcolor{orange!80!black}{\textsf{\footnotesize [CAVEAT: #1]}}}
\else
  \newcommand{\todo}[1]{}
  \newcommand{\note}[1]{}
  \newcommand{\caveat}[1]{}
\fi

\newcommand{\Uav}{\bar{U}}                       
\newcommand{\Dat}{\Delta_{\mathrm{at}}}          
\newcommand{\Uc}{U_c}
\newcommand{\JU}{J/U}

\newcommand{\Imut}{I(1\!:\!2)}

\newcommand{\wNS}{w_{N,S}}

\newcommand{\ket}[1]{|#1\rangle}

%% file: body.tex

\title{Hund physics in the quarter-filled two-band Hubbard--Kanamori model: Crystal-field splitting and non-local Correlations}

\author{Diego Florez-Ablan}
\affiliation{International School for Advanced Studies (SISSA), Via Bonomea 265, 34136 Trieste, Italy}

\author{Carlos Mejuto-Zaera}
\affiliation{Univ Toulouse, CNRS, Laboratoire de Physique Théorique, Toulouse, France.}

\author{Massimo Capone}
\affiliation{International School for Advanced Studies (SISSA), Via Bonomea 265, 34136 Trieste, Italy}
\affiliation{Istituto Officina dei Materiali (CNR-IOM), Via Bonomea 265, 34136 Trieste, Italy}

\date{\today}

\begin{abstract}

We address the effect of crystal-field splitting and non-local correlations in the quarter-filled two-orbital Hubbard–Kanamori model. The role of the Hund's coupling in this configuration is not obvious since the atomic ground state has only one electron per site. Yet, previous works identified fingerprints of the Hund's coupling defining a  weak Hund's metal at least at finite temperature.

Using single-site dynamical mean-field theory (DMFT) and two-site cellular DMFT based on a selected configuration-interaction solver we investigate the zero-temperature properties of the weak Hund's metal. Within DMFT we draw the boundary for the weak Hund's metal, which is consistent with previous finite-temperature results.

Interestingly, we find that the fingerprints of the Hund's coupling are quite robust to the lifting of the orbital degeneracy and they survive up to splittings of a quarter of the bandwidth, well beyond those at which a two-stage Kondo screening disappears.


Including non-local correlations in cluster DMFT demonstrates the robustness of the single-site DMFT picture, even if Mott localization occurs at smaller interactions. 
In the metal the intersite correlations are weak and insensitive to J, while in the localized state we find nearest-neighbor ferromagnetic, antiferro-orbital correlations that are manifest also in the mutual information between sites. 


\end{abstract}

\maketitle



\section{Introduction}
\label{sec:intro}

The exchange Hund's coupling $J$ is now recognized as a source of strong electronic correlation in multiorbital metals even far from a Mott insulator with distinctive properties that lead to introduce the name of Hund's metal~\cite{georgesStrongCorrelationsHunds2013,hauleCoherenceIncoherenceCrossover2009,wernerSpinFreezingTransition2008,georgesHundmetalPathStrong2024,Capone2026}.
One of the first characterizations of the Hund's metal has been the twofold effect of the Hund's exchange, a phenomenology coined \emph{Janus effect}~\cite{demediciJanusFacedInfluenceHunds2011,demediciHundsCouplingIts2011}: a reduction of the quasiparticle weight $Z$, i.e., of the electronic coherence, in the metallic states and an increase of the interaction needed to reach the Mott insulator. Its dynamical counterpart is spin--orbital separation, in which the spin degrees of freedom are screened at a much lower energy than the orbital ones~\cite{stadlerDynamicalMeanFieldTheory2015,georgesHundmetalPathStrong2024}. The paradigmatic example is the three-orbital model with two electrons, $(M,N)=(3,2)$, where $J$ splits the atomic ground-state manifold and builds a local $S=1$ moment, and the above phenomenology has been widely documented.

In this work we focus on a different configuration, a quarter-filled system with two orbitals with one electron per site, $(M,N)=(2,1)$. Here the atomic ground state is an $S=1/2$ orbital doublet that $J$ does not split. This led to the expectation that Hund physics is unimportant at quarter filling, a ``conventional wisdom'' that Ryee \textit{et al.}~\cite{ryeeHundPhysicsLandscape2021} revisited: in single-site dynamical mean-field theory (DMFT), the $(2,1)$ case shows Hund-metal behavior once the interaction is measured by its average, $U-J$ (see below for the definition of the Hamiltonian).
Kugler \textit{et al.}~\cite{kuglerLowenergyPerspectiveTwoorbital2024} confirmed with a numerical renormalization-group (NRG) solver that a two-stage Kondo screening (2SKS), reflecting a kind of splitting in the local degrees of freedom, is present also at $(2,1)$, while the Janus effect and spin--orbital separation are present but weak. Earlier DMFT work on the same model addressed the Mott transition~\cite{kogaMottTransitionsTwoOrbital2005,inabaMetalinsulatorTransitionTwoorbital2007}, ferromagnetism and orbital order~\cite{heldMicroscopicConditionsFavoring1998,petersOrbitalMagneticOrder2010}, and the charge-disproportionated regime at $J/U>1/3$~\cite{isidoriChargeDisproportionationMixed2019}.

The question about this non-obvious Hund state is indeed directly relevant to some materials. In the infinite-layer nickelates $R$NiO$_2$, the debate is focused on the alternative between an effective one-band description ~\cite{kitataniNickelateSuperconductorsRenaissance2020,siClosingPossibleScenarios2024} and a Hund's metal picture involving the two Ni $e_g$ orbitals~\cite{wangHundsMetalPhysics2020,kangOpticalPropertiesInfiniteLayer2021,kangInfinitelayerNickelatesNieg2023,lechermannMultiorbitalProcessesRule2020} with a filling of about one hole per Ni.
In this context, two-orbital low-energy models have been constructed from first principle calculations~\cite{PhysRevLett.125.077003,kuglerLowenergyPerspectiveTwoorbital2024} and, in all of these, the $e_g$ orbitals are subject to an additional term, a crystal field splitting $\delta$. 
This parameter is known to compete with the Hund's coupling, favoring orbital differentiation and low-spin states, and suppressing 2SKS~\cite{kangInfinitelayerNickelatesNieg2023,kuglerLowenergyPerspectiveTwoorbital2024}.
Moreover, the size of $\delta$ depends on how the low-energy model is built: estimates range from a splitting well below $U$ and $J$~\cite{kangInfinitelayerNickelatesNieg2023} to the order of the bandwidth~\cite{lechermannMultiorbitalProcessesRule2020,kuglerLowenergyPerspectiveTwoorbital2024}, and correlations renormalize it further~\cite{kangOpticalPropertiesInfiniteLayer2021}. The relevance of two-orbital Hund metal physics for the Nickelates therefore hinges on two questions: how large $\delta$ is, and how robust Hund physics is against it. 


Kugler \textit{et al.}~\cite{kuglerLowenergyPerspectiveTwoorbital2024} showed that 2SKS at $(2,1)$ is fragile: on a Bethe lattice of bandwidth $W$, at $U=1.55W$ and $J/U\approx0.23$, it is marginal at $\delta=0.04W$ and absent at $\delta=0.11W$, in line with the criterion that 2SKS requires $|\delta|\lesssim T_K^{\mathrm{orb}}(\delta=0)/2$, with $T_K^{\mathrm{orb}}(\delta=0)$ the orbital Kondo scale of the degenerate model. Spin--orbital separation persists at larger $\delta$, but it originates from atomic energies rather than from 2SKS. 

However, at least two questions need to be answered to settle the problem. 
First, how do the other Hund signatures, such as the $J$ dependence of $Z$ and of the high-spin multiplet weight, respond to a crystal field across $J/U$.
Second, as both Refs.~\cite{ryeeHundPhysicsLandscape2021,kuglerLowenergyPerspectiveTwoorbital2024} use single-site DMFT, the effect of non-local correlations is still unexplored.

Only a few calculations have addressed non-local effects in Hund's metal, but they did not discuss these aspects. Kita \textit{et al.}~\cite{kitaSpatialFluctuationsSpin2009} studied the same quarter-filled two-orbital model on the square lattice with two-site cellular DMFT and a non-crossing-approximation solver at temperatures $T\geq0.03t$. They found that $J$ enhances antiferro-orbital fluctuations and suppresses the nearest-neighbour antiferromagnetic spin correlations at low temperature, driving them toward a ferromagnetic sign, and reported a pseudogap at $U=W/2$.
Nomura \textit{et al.} ~\cite{nomuraNonlocalCorrelationsInduced2015} studied Hund-induced nonlocal correlations in the three-orbital model at the $(3,2)$ filling at finite temperature with a two-patch dynamical cluster approximation, where they found an anomalous metallic phase at high-U and finite J with ferromagnetic inter-site spin correlations. They found a momentum-space differentiation
with an insulating, half-filled $\Gamma$ patch and a coherent $M$ patch. 
However, whether the Hund metal at quarter filling develops comparable J-driven non-local correlations at T = 0, and how these local Hund's metal signatures are affected by short-range correlations has not been addressed.

Here we address the two above mentioned questions at zero temperature, combining single-site DMFT and two-site cellular DMFT (CDMFT) with a selected configuration-interaction impurity solver that retains the full Kanamori interaction, including spin-flip and pair-hopping terms and crystal-field splitting.
For degenerate bands we follow the $J$ dependence of the quasiparticle weight over the whole range $0\le J/U<1/3$, holding fixed in turn the bare interaction $U$, the average interaction $U-J$, and the lowest atomic charge gap $U-3J$ to better highlight all the different aspects of the weak Hund metal.
We then include a crystal field splitting and we resolve these responses by orbital, for splittings $\delta=t$ and
$2t$ ($0.125W$ and $0.25W$), comparable to or larger than the splitting at which 2SKS has disappeared~\cite{kuglerLowenergyPerspectiveTwoorbital2024}.
Finally, we include the minimal short-range correction, a two-site cluster, and use its reduced density matrices to separate local from non-local correlations.

Section~\ref{sec:model} introduces the model, the method and the atomic limit. Secs.~\ref{sec:delta_0} and~\ref{sec:cfs} present the single-site results without and with crystal field, and sec.~\ref{sec:cluster} discusses the two-site results.

\section{Model and method}
\label{sec:model}

We consider the two-band Hubbard--Kanamori model ($M=2$) on the square lattice, with band-diagonal nearest-neighbor hopping $t$ (bandwidth $W=8t$), at quarter-filling (one electron per site):

\begin{align}
H &= -t\sum_{\langle ij \rangle m \sigma} \left( d^{\dagger}_{i m \sigma} d_{j m \sigma} + h.c. \right) + \sum_{im\sigma}  \left(\delta_m - \mu\right)n_{im\sigma}
   \\&+ \sum_i H_{\mathrm{int}}[d_{im\sigma}] .
\label{eq:H}
\end{align}
where $\langle ij \rangle$ denotes the sum over nearest-neighbor pairs, $\mu$ is the chemical potential  and $\delta_m = \mp \delta/2$ for $m=1,2$ is the crystal field splitting, such that for $\delta>0$ band 1 is lower in energy than band 2. Both bands have the same dispersion, $\epsilon_k = -2t (\cos{k_x}+\cos{k_y})$, and become degenerate at $\delta=0$.
The interaction term is given by the standard form of the two-band Hubbard--Kanamori interaction, including both the spin-flip and pair-hopping terms, which can be written as

\begin{align}
H_{\mathrm{int}} &= \frac{U-3J}{2}N(N-1) - 2J\bm{S}^{2}
\nonumber\\ &\quad
- 2J\left(T_\text{band}^y\right)^{2} + 2JN ,
\label{eq:Hint_HK}
\end{align}
where $N=\sum_{m\sigma}n_{m\sigma}$ is the local particle number,
$\bm{S}$ the total local spin, and $T_\text{band}^y=-\frac{i}{2}\sum_\sigma\left(d^{\dagger}_{1\sigma} d_{2\sigma}-d^{\dagger}_{2\sigma} d_{1\sigma}\right)$ denotes the $y$-component of the orbital pseudo-spin.
The total orbital pseudo-spin is given by $\bm{T}_{\text{band}}=\frac{1}{2}\sum_{mm'\sigma}d^{\dagger}_{m\sigma} \bm{\sigma}_{mm'} d_{m'\sigma}$, with $\bm{\sigma}$ the vector of Pauli matrices. The $y$-component of the orbital pseudo-spin also satisfies the  relation $T_\text{band}^y = \frac{L^z}{2}$, where $L^z$ is the $z$-component of the orbital angular momentum operator.

In order to highlight different aspects of the Hund's physics \cite{ryeeHundPhysicsLandscape2021,kuglerLowenergyPerspectiveTwoorbital2024}, we study the dependence on $J$ while holding fixed one of the three interaction scales: 
\begin{align*}
U , \qquad&
\Uav = U - J , \\
&\Dat = U - 3J ,
\label{eq:gauges}
\end{align*}
identified as the intra-band pair cost, the average of the orbital interaction tensor, and the atomic charge gap respectively.


We solve the model with single-site DMFT~\cite{georgesDynamicalMeanfieldTheory1996} and its two-site cellular extension (CDMFT)~\cite{kotliarCellularDynamicalMean2001,maierQuantumClusterTheories2005}, retaining both orbitals on each site\cite{nomuraMultiorbitalClusterDynamical2014}.
The two-site cluster contains one nearest-neighbour bond, so that each site has one of its four neighbors inside the cluster, and the other three bonds enter only through the self-consistent bath.

All calculations are at $T=0$, and we restrict to paramagnetic solutions. The impurity  problem is represented by a Hamiltonian with a discrete bath, whose ground state and Green's function are computed with the adaptive sampling configuration interaction (ASCI) solver ~\cite{tubmanModernApproachesExact2020,williams-youngParallelDistributedMemory2023}, following its use within DMFT \cite{mejuto-zaeraDynamicalMeanField2019,bath_yo}.
ASCI builds a truncated many-body basis by iteratively selecting the Slater determinants with the largest estimated contributions to the ground state. The ground-state expansion is limited to $N_\mathrm{dets}\sim 10^5-10^6$ determinants depending on the parameter point.

The DMFT self-consistency cycle is carried out on the imaginary-frequency axis using $N_\omega=1000$ positive fermionic Matsubara frequencies, $\omega_n=(2n+1)\pi/\beta$, with $\beta t=157$ and $\omega_{\max}\simeq40t$. Here $\beta$ defines an auxiliary frequency grid and does not represent the physical temperature of the ground-state calculation.
At each iteration, the hybridization function obtained from the self-consistency condition is fitted by a discrete bath consisting of $N_b$ spin-degenerate bath orbitals. We use $N_b=14$ bath orbitals for the single-site DMFT calculations and $N_b=12$ bath orbitals for the two-site cluster calculations.
We use two different parametrizations for the bath: the semidefinite-relaxation (SDR) parametrization \cite{mejuto-zaeraEfficientHybridizationFitting2020} and a symmetry-adapted parametrization based on the irreducible representations of the cluster \cite{kochSumRulesBath2008,senechalBathOptimizationCellular2010,bath_yo}. 
The former uses a bath representation suited to semidefinite relaxation of the fitting problem, while the latter organizes the bath couplings into cluster symmetry channels.

We estimate the quasiparticle residue from the diagonal self-energy at the lowest Matsubara frequency,
\begin{equation}
Z_{im} =
\left[
1-\frac{\operatorname{Im}\Sigma_{im,im}(i\omega_0)}
{\omega_0}
\right]^{-1}
\end{equation}
where $i$ labels the cluster site (absent in single-site DMFT) and $m$ labels the band.
This finite-frequency estimate approximates the zero-frequency self-energy slope in the metallic regime. 

In a gapped state the slope estimate need not vanish and is not a quasiparticle residue for the case of cluster embedding. We therefore classify a two-site point as insulating whenever the imaginary part of the local Green's function is below a certain threshold, and display $Z_i = 0$ there.


\subsubsection{Atomic limit of the model}
\label{sec:atomic}

We can gain some insight on our model from an analysis of the local atomic Hamiltonian of the two-band degenerate model ($\delta = 0$). The energies of the $N=0$ and $N=1$ charge sectors are simply $E(0)=E(1)=0$, while the six states in the $N=2$ sector are split into three groups:
\begin{align}
  E(2) = \begin{cases}
  U-3J, & \text{for } \ket{S=1,T_{\text{band}}=0,T^y_{\text{band}}=0}, \\
  U-J, & \text{for } \ket{S=0,T_{\text{band}}=1,T^y_{\text{band}}=\pm1}, \\
  U+J, & \text{for } \ket{S=0,T_{\text{band}}=1,T^y_{\text{band}}=0}.
  \end{cases}
\label{eq:E2multiplets}
\end{align}
These are the three spin-triplet states at lower energy, and the spin-singlet states which are split according to their orbital character, with the symmetric combination of intra-band doublons being the highest energy state.
The charge gap is then given by
\begin{equation}
\Dat \equiv E(2)+E(0)-2E(1)=U-3J .
\label{eq:atomicgap}
\end{equation}

For all $J/U<1/3$, the atomic ground state at $\langle N \rangle=1$ is therefore an $S=1/2$ orbital doublet (4-fold degenerate), which is not affected by a finite value of $J$.
The main effect of $J$ in this limit is to reduce the atomic charge gap $\Dat$ from the singly occupied state to the inter-band triplet. At $J/U=1/3$, the charge gap closes, and the ground state becomes an eight-fold degenerate manifold of $N=0$, $N=1$, and $N=2,S=1$ states. Finally, for $J/U>1/3$, the charge gap becomes negative, and the atom is a degenerate mixture of empty and high-spin half-filled states\cite{isidoriChargeDisproportionationMixed2019}.

This atomic structure is what distinguishes $(M,N)=(2,1)$ from the three-band case at $N=2$. In that case, $J$ splits the atomic ground state manifold (15-fold down to 9-fold), building a local $S=1$ moment and producing the conventional Janus effect
\cite{demediciJanusFacedInfluenceHunds2011,ryeeHundPhysicsLandscape2021}.
That route is genuinely absent here. This is the ``conventional wisdom'' that led to the disregard of Hund's physics at quarter filling Ref.~\cite{ryeeHundPhysicsLandscape2021}.
However, it turned out that the Hund's coupling can affect the low-energy physics even without a splitting of the atomic ground state: 
splitting the $N=2$ manifold is already enough to change the low-energy physics of the system. 
The Kondo couplings obtained by the Schrieffer-Wolff transformation scale as $\mathcal{J}^i\sim V^2/\Delta E$ with $\Delta E$ the charge-excitation energy. Lowering the inter-band triplet to $U-3J$ enhances the virtual fluctuation between $\ket{N{=}2,S{=}1}$ and $\ket{N{=}1,S{=}1/2}$, suppressing the antiferromagnetic Kondo coupling $\mathcal{J}^{\mathrm{spin}}$ relative to the orbital one~\cite{ryeeHundPhysicsLandscape2021,stadlerDynamicalMeanFieldTheory2015}.  
As for $M\ge3$, $J$ thus suppresses spin screening, but here entirely through excited states.
This mechanism is general, the orbital-singlet, maximal-spin intermediate state of the two-stage Kondo screening is available whenever the impurity is one electron away from half-filling ($N=M-1$).
Nevertheless, the ground state mechanism being absent in $(M,N)=(2,1)$, the resulting Hund-metal signatures are weaker in this regime than those of the conventional $(M,N)=(3,2)$ case \cite{ryeeHundPhysicsLandscape2021,kuglerLowenergyPerspectiveTwoorbital2024}.

However, the question of the robustness of this physics to crystal field remains crucial, since, in realistic systems, 
the crystal field splitting is always present to some extent \cite{kangInfinitelayerNickelatesNieg2023}.
Adding such a crystal field changes the atomic balance. A term $\delta_{1,2}=\mp\delta/2$ splits the $N=1$ states with energies $E(N=1)=\pm|\delta|/2$,  while pair hopping mixes the two intra-band doublons to produce singlet states at $E_{N=2,S=0,T_\text{band}=1}=U\pm\sqrt{J^2+\delta^2}$. In this symmetric split setting, the inter-band triplet and singlet states remain at the same energies.
The atomic charge gap must therefore be formed with the lower state between the triplet and new low-energy singlet, giving:
\begin{equation}
\Dat(\delta)
=
\begin{cases}
U+|\delta|-3J,
& \displaystyle \frac{J}{|\delta|}\geq\frac{\sqrt{2}}{4}
,\\[8pt]
U+|\delta|-\sqrt{J^2+\delta^2},
& \displaystyle \frac{J}{|\delta|}\leq\frac{\sqrt{2}}{4}
.
\end{cases}
\label{eq:atomicgap_delta}
\end{equation}
The two branches join continuously at
$J/|\delta|=\sqrt{2}/4$, where singlet and triplet are degenerate.
The crystal field thus stabilizes the singly occupied configuration by increasing the charge gap, and it favors low-spin over high-spin doublons in the $N=2$ sector.

A finite $\delta$ also lifts the band degeneracy that the Kondo physics relies on, and cuts off band fluctuations below the energy scale $\delta$. 
As a result, two-stage Kondo screening is suppressed once the splitting competes with the orbital Kondo scale $|\delta|\gtrsim T_K^{\mathrm{orb}}(\delta=0)/2\,$\cite{kuglerLowenergyPerspectiveTwoorbital2024}.



\section{Degenerate bands (\texorpdfstring{$\delta=0$}{δ=0})}
\label{sec:delta_0}

We first discuss the Hund signatures of the orbital-degenerate model at quarter filling within our T=0 DMFT which represents a reference to include the crystal-field and two-site correlations, and at the same time a validation of our approach against the literature.

\begin{figure}[tb]
  \centering
  \includegraphics[width=\linewidth]{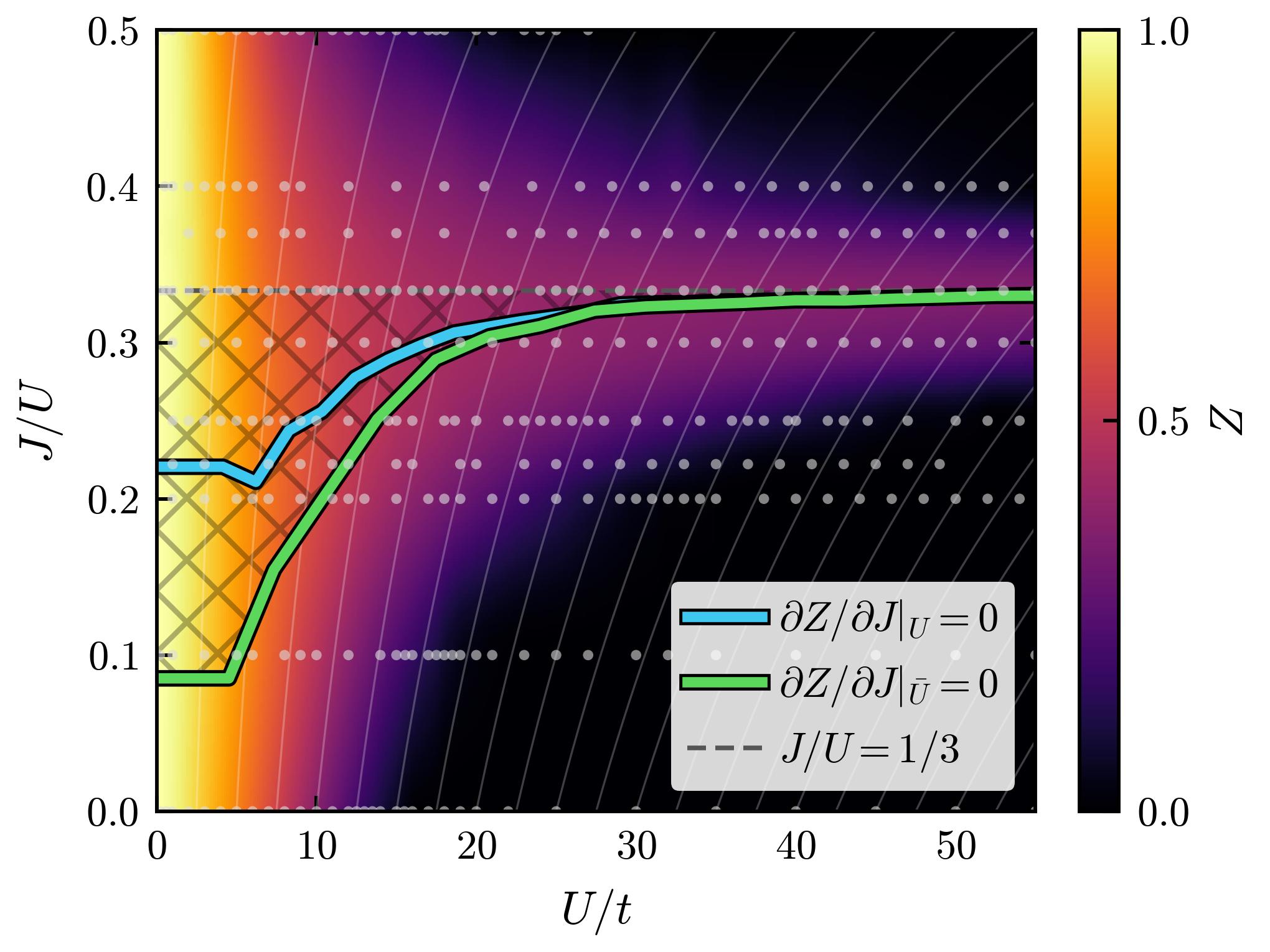}
  \caption{
  Quasiparticle weight $Z$ of single-site DMFT at $\delta=0$ in the $(U/t,\JU)$ plane. Grey dots are the DMFT data; the map interpolates between them. Thin curves are contours of constant $\Uav$.
  The thick lines mark the maximum of $Z$ along cuts at fixed $U$ (cyan, $\partial Z/\partial J|_U=0$) and at fixed $\Uav$ (green, $\partial Z/\partial J|_{\Uav}=0$). 
  Above each line the corresponding derivative is negative.}
  \label{fig:zlandscape}
\end{figure}

\begin{figure*}[tb]
  \centering
  \includegraphics[width=0.98\textwidth]{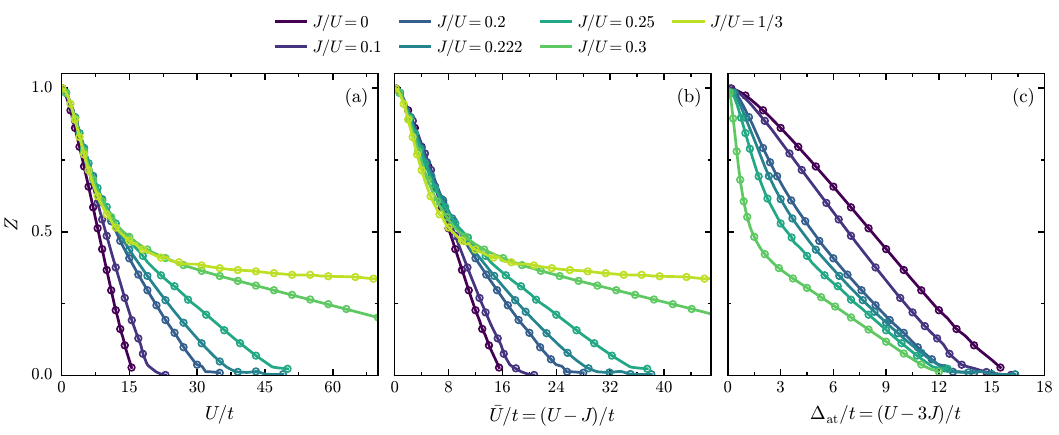}
  \caption{Quasiparticle weight at $\delta=0$ against (a) the bare $U$, (b) $\Uav=U-J$, and (c) $\Dat=U-3J$. $\JU \geq 1/3$ is excluded from (c) since $\Dat\leq 0$ there.}
  \label{fig:gauges}
\end{figure*}

The Janus criterion in the form used in Ref.~\cite{ryeeHundPhysicsLandscape2021} is the conjunction of the two conditions:
\begin{equation}
\frac{d\Uc(J)}{dJ} > 0 
\qquad\text{and}\qquad
\left(\frac{\partial Z}{\partial J}\right)_{\mathrm{X}} < 0.
\label{eq:janus}
\end{equation}
where $X$ is one of the three interaction scales $U$, $\Uav$, or $\Dat$.

In figure ~\ref{fig:zlandscape}, we show the quasiparticle weight $Z$ in the $(U/t, J/U)$ plane at $\delta=0$. 
There, we interpolated the $Z$ data to produce a smooth landscape and compute the derivatives $\partial Z/\partial J|_{U}$ and $\partial Z/\partial J|_{\Uav}$. 
The colored lines shown in the figure correspond to the points where these derivatives vanish, separating the regions where $Z$ increases or decreases with $J$ at fixed $U$ or $\Uav$. 
Above the lines, the derivative is negative, indicating that increasing $J$ suppresses the quasiparticle weight.

At fixed U (Fig. ~\ref{fig:zlandscape}), $U_c$ increases strongly with $J$, so the first condition of Eq.~\eqref{eq:janus} is satisfied throughout the physical range $0<\JU<1/3$.
However, focusing on the traditional marker for the Janus effect $\partial Z/\partial J|_{U}$ (cyan line in Fig.~\ref{fig:zlandscape} and Fig.~\ref{fig:gauges}(a)), we see that the effect is very small and present only for $J/U$ close to the $1/3$ limit. The trend is dominated by the atomic gap effect: 
increasing $J$ makes $\Dat$ shrink fast, and the quasiparticle weight $Z$ increases because the atomic gap closes.

Holding $\Uav$ fixed instead (green line in Fig.~\ref{fig:zlandscape} and Fig.~\ref{fig:gauges}(b)) removes part of this shift.
The $\partial Z / \partial J|_{\Uav} < 0$ region is much broader. We thus recover the weak-Hund region introduced in Ref.~\cite{ryeeHundPhysicsLandscape2021}, where a Bethe lattice was used instead of our square lattice.

In Fig.~\ref{fig:gauges}(c), we keep the atomic gap $\Dat$ (Eq. \eqref{eq:atomicgap}) in order to test whether the only effect of $J/U$ in the degree of correlation is due to atomic effects. If that were the case, we would expect the curves to collapse.
Instead, throughout their common metallic range, we observe that $Z$ at fixed $\Dat$ decreases with $\JU$, and the $J=0$ curve is the least correlated, i.e. $Z(\Dat)|_{J\neq0}<Z(\Dat)|_{J=0}$. 
This comparison also matches the results from Ref.~\cite{kuglerLowenergyPerspectiveTwoorbital2024} and confirms that the effect of $J/U$ on the degree of correlation is not controlled only by atomic physics.

\begin{figure*}[tb]
  \centering
  \includegraphics[width=0.6\textwidth]{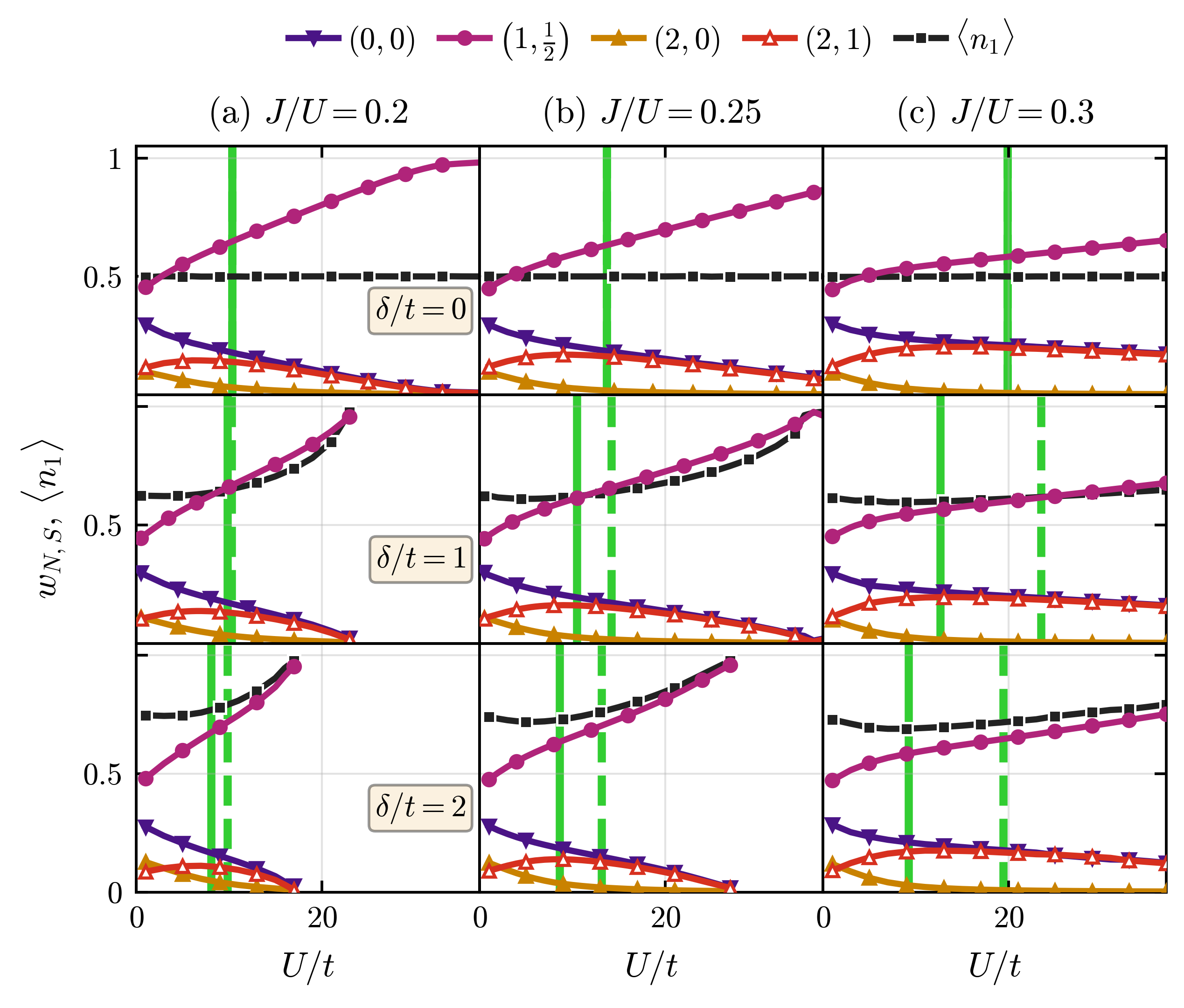}
  \caption{Projection into the local atomic multiplets of the impurity's reduced density matrix $\wNS=\mathrm{Tr}(\rho P_{N,S})$ for different values of $\JU$ and $\delta$ inside the weak Janus effect region. The vertical lines 
  mark the values that delimit the weak Janus regime, i.e., the points where $dZ_m/dJ|_{\Uav}=0$ for each $\JU$.}
  \label{fig:multiplets}
\end{figure*}

We now move to an inspection of the local atomic multiplets, reported in Fig. ~\ref{fig:multiplets}. These are obtained projecting the converged impurity reduced density matrix onto the local atomic multiplets $\wNS=\mathrm{Tr}(\rho P_{N,S})$, where $P_{N,S}$ is the projector into the $(N,S)$ multiplet (see App. \ref{app:rdm}), and we show the results for $\JU=0.2$ and $\JU=0.25$ inside the weak Hund region. The results for the orbital-degenerate case are shown in the first row of the panel. 
We observe how the Hund's coupling splits the $N=2$ manifold, favoring the inter-band triplet over the singlet states in the atomic decomposition of the state. 
This delays the recovery of the pure $N=1,S=1/2$ atomic ground state limit and enhances $\langle(S^z)^2\rangle$ above its single spin value of 1/4. Thus, even though Hund coupling does not split the quarter-filled atomic ground state, it reshapes the virtual charge fluctuations that control the low-energy physics.
Moreover, we have added as a visual guide the values of $U/t$ such that $dZ/dJ|_{\Uav}=0$. For the values of $U/t$ below this point, the weak Janus effect is present, and we can see that this region coincides with the region of increasing $(N=2,S=1)$ weight in the atomic decomposition.
This shows that the two main static local markers for Hundness, the Janus effect and the presence of increasingly important high spin weight in the multiplet decomposition of the ground state coincide. As the system leaves the Janus effect region, the high spin contribution also starts decreasing.

\section{Crystal-field splitting}
\label{sec:cfs}

\begin{figure*}[tb]
  \centering
  \includegraphics[width=\textwidth]{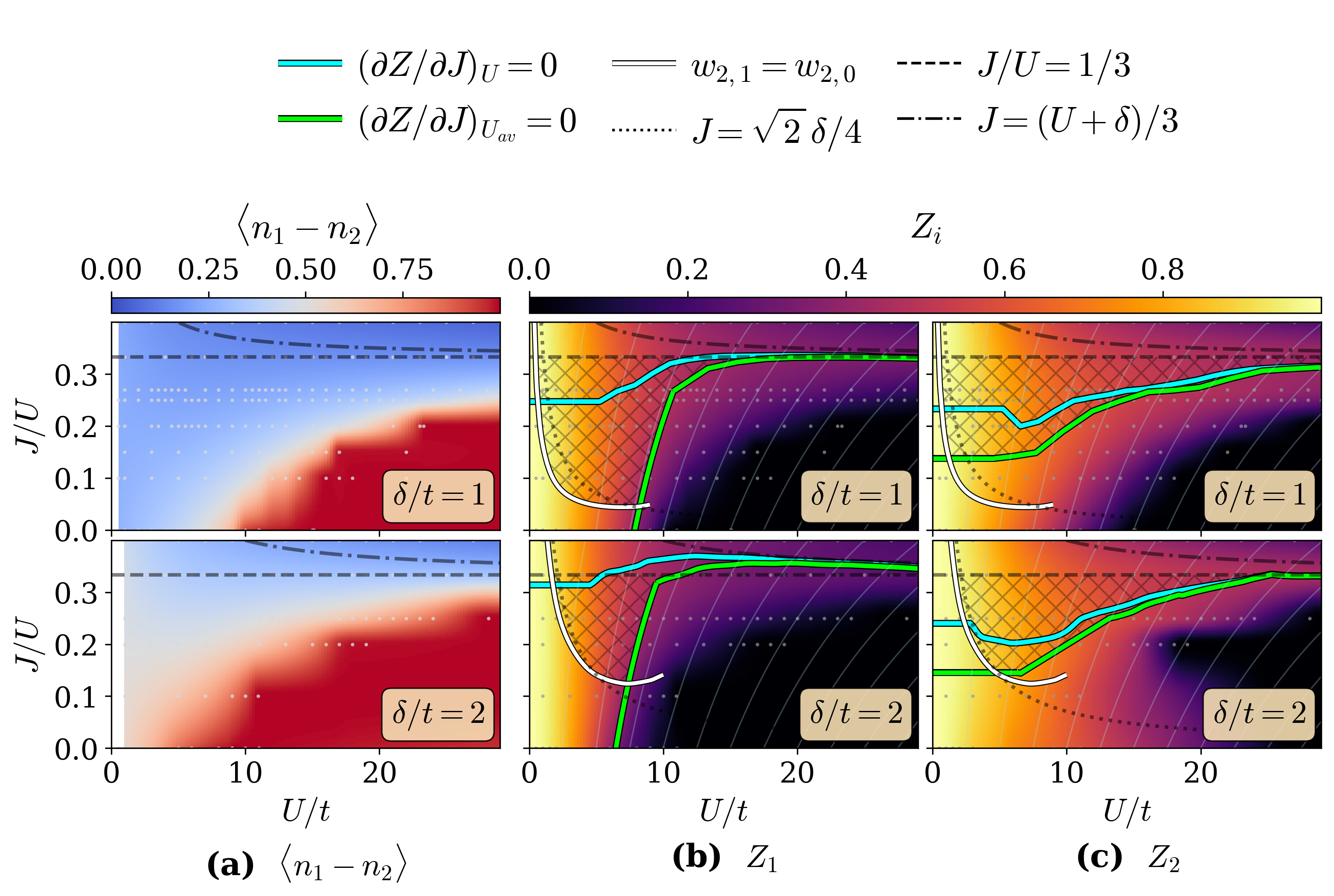}
  \caption{ Dependence of the weak Hund's metal region with crystal field splitting. 
  Crystal-field dependence of  the band polarization $\langle n_1-n_2\rangle$ (a), and  the quasiparticle weights $Z_1$ (b) and $Z_2$ (c) in the $(U/t,\JU)$ plane. 
  The top and bottom rows correspond to $\delta/t=1$ and $2$, respectively. Cyan and green solid curves mark $\partial Z_m/\partial J|_U=0$ and $\partial Z_m/\partial J|_{\Uav}=0$ for the corresponding band. White curves mark equal triplet and singlet weights, $w_{2,1}=w_{2,0}$. 
  The dotted curve indicates the atomic crossing $J=\sqrt{2}\delta/4$, and the dashed and dot-dashed lines mark the points where $\Delta_{at}=0$ at $\delta=0$ ($\JU=1/3$) and $\delta>0$ ($J=(U+\delta)/3$).}
  \label{fig:doping_landscapes}
\end{figure*}

In this section we focus on the central question of the role of the crystal-field splitting on the weak Hund picture that we described above. A finite crystal-field splitting lifts the band degeneracy so that the two bands have different properties, including quasiparticle weights. Fig.~\ref{fig:doping_landscapes} shows the resulting phase diagrams for $\delta/t=1$ and $\delta/t=2$.
At large $U/t$ and small  $J/U$, the interaction amplifies the crystal-field-induced orbital polarization $n_1-n_2$ and drives the system toward the band polarized atomic limit, $\langle n_1-n_2\rangle\to1$, in which band 1 is half-filled and band 2 is empty.
However, the Hund's coupling competes with this tendency. $J$ lowers the energy of the inter-band triplet and hence promotes the occupation of both bands. Increasing $J$ at fixed interaction strength therefore reduces the band polarization.

This effect of $J$ on the occupations has direct consequences for the quasiparticle weights as in the simplest scenario, the $Z$ decreases when the occupation of an orbital approaches half-filling, while the doping away from half-filling leads to a larger $Z$.

Thus, $J$ influences the quasiparticle weights both directly, through the Hund-induced suppression of coherence, and indirectly, through the $J$-dependent redistribution of charge between the bands.

This redistribution of charge has opposite consequences for the two quasiparticle weights. For band 1, the decrease of $\langle n_1\rangle$ moves the band away from integer filling and tends to increase $Z_1$, opposing the intrinsic Hund-induced suppression of coherence. For band 2, by contrast, the increase of $\langle n_2\rangle$ tends to reduce $Z_2$ and thus reinforces the direct Hund-induced suppression. 

The weak Janus effect is consequently strongly reduced for the more occupied band, where the two mechanisms compete, but remains clearly visible for the less occupied band, where they act in the same direction. This also explains why the two ridges marking the zero derivative loci of $J$ separate once $\delta\neq0$. The region in which the more occupied band displays a negative $\left.dZ_1/dJ\right|_{\Uav}$ is significantly reduced as $\delta$ increases, while the less occupied band retains a negative $\left.dZ_2/dJ\right|_{\Uav}$ over a similar range of $U/t$ as the degenerate case. 

The multiplet decomposition provides an interpretation of this competition.   The second and third rows of Fig.~\ref{fig:multiplets} show the results for $\delta/t=1$ and $2$, respectively. As $U/t$ increases, charge fluctuations are suppressed, and the $N=1$ sector becomes dominant. Before this atomic limit is reached, however, the Hund's coupling transfers weight within the $N=2$ sector from the singlet configurations to the inter-orbital $(N=2,S=1)$ triplet. The triplet is favored over the singlets over a broader interval than what the level crossings of the isolated atom would suggest (eq. \eqref{eq:atomicgap_delta}). Correspondingly, the Hund's coupling delays the approach to the fully band polarized state visible in Fig.~\ref{fig:doping_landscapes}(a).

The vertical lines in Fig.~\ref{fig:multiplets} indicate the values of $U/t$ at which $\left.dZ_m/dJ\right|_{\Uav}$ changes sign. As in the degenerate case, the region in which the weight of the triplet grows with interaction coincides with the weak-Janus regime. This correlation supports the interpretation that the remaining suppression of $Z_m$ is not solely a consequence of the $J$-induced transfer of the occupations, and it is also still tied to the enhanced high-spin character of the virtual $N=2$ fluctuations.

Remarkably, a finite region in which both bands display a negative $\left.dZ_m/dJ\right|_{\Uav}$ survives even for $\delta/t=2$, together with an enhanced triplet weight. Thus, although the crystal field suppresses band fluctuations and strongly weakens the Janus response of the more occupied band, it does not immediately eliminate all the signatures of the Hund metal.

The above analysis has shown that the local signatures of Hund's coupling are much less sensitive to the crystal field than the reported effect on the two-stage Kondo screening. For $U=3.1D$ and $J=0.7D$ on the Bethe lattice, i.e.\ $U\approx12.4t$ and $J/U\approx0.23$ at equal bandwidth, Ref.~\cite{kuglerLowenergyPerspectiveTwoorbital2024} finds two-stage screening weakened at $\Delta=0.08D\approx0.3t$ and absent at $\Delta=0.22D\approx0.9t$. By contrast, three signatures persist at $\delta=t$ and $2t$: a negative $(\partial Z_m/\partial J)_{\Uav}$ in the less occupied band, a dominant triplet weight, and the $J$-driven reduction of the polarization. The difference reflects the driving energy scales for each phenomenon. Two-stage screening requires orbital fluctuations down to the orbital Kondo scale and it is lost once $|\delta|\gtrsim T_K^{\rm orb}(0)/2$~\cite{kuglerLowenergyPerspectiveTwoorbital2024}, whereas the local signatures are controlled by atomic energies, $J$ against $\delta$ and $\Dat$, which are much larger. For the less occupied band the $J$-driven charge transfer contributes to the persistence, so these signatures need not have the same microscopic origin as at $\delta=0$. The two-orbital model of LaNiO$_2$ in Ref.~\cite{kuglerLowenergyPerspectiveTwoorbital2024} shows the same contrast: with $\Delta\approx0.8$\,eV two-stage screening is absent, yet at $J=0.7$\,eV the triplet weight exceeds the singlet weight and $J$ reduces the polarization.

\section{Non-local correlations via two-site cellular DMFT}
\label{sec:cluster}



In this section we explore the effect of non-local correlations on the weak Hund state using cluster DMFT with a two-site cluster at $\delta=0$. As discussed in Sec.~\ref{sec:model}, our approach amounts to singling out one of the four bonds connecting a site with the rest of the lattice and treating it exactly, while the remaining bonds are effectively described by the presence of the self-consistent bath. It should be clear that the bond included in the cluster can be overweighted in this procedure~\cite{Capone1d,sakaiClustersizeDependenceCellular2012}. 

In this section we discuss whether the local Hund signatures of Sec.~\ref{sec:delta_0} survive this correction, and which correlations the explicit bond adds.

We characterize the solution via a variety of quantities. Besides  the on-site quasiparticle weight $Z_i$ (Sec.~\ref{sec:model}), we consider reduced density matrices (App.~\ref{app:rdm}): the cluster density matrix $\rho_{12}$, obtained by tracing the bath out of the impurity ground state, and the local $\rho_i$, obtained by also tracing out the other cluster site. The multiplet weights of a cluster site, $w^{(i)}_{N,S}=\mathrm{Tr}(\rho_iP_{N,S})$, are the counterpart of the single-site $\wNS$, with $w^{(1)}=w^{(2)}$ by symmetry.
We also compute orbital-resolved density-density correlators
\begin{equation}
C^{ij}_{mm'}=\langle n_{im}n_{jm'}\rangle-\langle n_{im}\rangle\langle n_{jm'}\rangle ,
\qquad
n_{im}=\sum_\sigma n_{im\sigma} ,
\label{eq:Cij}
\end{equation}
which are local for $i=j$ and intersite for $i\neq j$. We write $C^{ij}_{mm}$ for the intra-band and $C^{ij}_{mm'}$ ($m\neq m'$) for the inter-band correlators; at $\delta=0$ they do not depend on the band labels. Since $\langle n_{im}\rangle=1/2$, $C^{ij}_{mm'}>0$ ($<0$) means that the charge fluctuations of $(i,m)$ and $(j,m')$ are correlated (anticorrelated).
The intersite correlators also measure the orbital correlations of the bond. With the orbital pseudospin of site $i$, $\tau^z_i=(n_{i1}-n_{i2})/2$, one has at $\delta=0$
\begin{equation}
\langle\tau^z_i\tau^z_j\rangle=\frac{1}{2}\left(C^{ij}_{mm}-C^{ij}_{mm'}\right),
\label{eq:tautau}
\end{equation}
so that $C^{ij}_{mm}<0<C^{ij}_{mm'}$ signals antiferro-orbital correlations. 
We also discuss the local magnetic moment $\langle(S^z_i)^2\rangle$ and the intersite correlator $\langle S^z_iS^z_j\rangle$, with $S^z_i=\frac{1}{2}\sum_m(n_{im\uparrow}-n_{im\downarrow})$.
Finally, we measure the total correlation between the two sites by the mutual information
\begin{equation}
\Imut=S(\rho_1)+S(\rho_2)-S(\rho_{12}) ,
\label{eq:mutinfo}
\end{equation}
where $S(\rho)=-\mathrm{Tr}\,\rho\ln\rho$ is the von Neumann entropy; both quantities are measured in units of $\ln 2$ (bits). $\Imut$ counts classical and quantum correlations together, not entanglement alone~\cite{bellomiaQuasilocalEntanglementMottHubbard2024}, and vanishes if and only if $\rho_{12}=\rho_1\otimes\rho_2$.

We first evaluate these quantities for an isolated dimer (Sec.~\ref{sec:dimer}), which provides a reference for the  large-$U$ limit, when the two-site cluster decouples from the bath. We then compare $Z_i$ and $w^{(i)}_{N,S}$ with the single-site results (Sec.~\ref{sec:cluster-local}), and finally follow the correlators and $\Imut$ as a function of the interaction (Sec.~\ref{sec:correlators}).


\subsection{Isolated two-site cluster}
\label{sec:dimer}

To interpret the results of two-site cellular DMFT at large $U$, we consider an isolated dimer with two bands per site hosting two electrons, and we start from $\delta=0$. Each two-electron state can be classified by its symmetry in the spin, band and site labels.
We find it useful to introduce the following complete set of commuting operators, which can be used to classify the states of the dimer Hilbert space in the quarter-filled sector:
\begin{equation}
\{N, S^2, S^z, T_{\text{band}}^2, T^{y}_\text{band}, T_{\text{site}}^2, T_{\text{site}}^z\} ,
\end{equation}
besides the standard spin and orbital pseudospin, we have introduced site pseudospin operators
$T_{\text{site}}^\alpha = \tfrac12\sum_{i,j=1,2}\sum_{m=1,2} \sum_{\sigma=\uparrow,\downarrow} d^\dagger_{im\sigma} \bm{\sigma}^\alpha_{ij} d_{jm\sigma}$.

The electronic states can be symmetric (triplet) or antisymmetric (singlet) under exchange in each of these three labels. Since the full state must be antisymmetric, the number of singlet factors must be odd: either all three labels are singlets, or exactly one is. This gives four blocks labeled by the total spin/pseudospin $\ket{S,T_{\text{band}},T_{\text{site}}}$:
\begin{equation}
\ket{0,0,0}, \quad
\ket{1,1,0}, \quad
\ket{1,0,1}, \quad
\ket{0,1,1},
\end{equation}
with one state in the first class and nine in each of the others
(Indeed $1+3\times9=28=\binom{8}{2}$, i.e., the size of the full dimer Hilbert space for two electrons).
Now, inside the dimer we can write the non-interacting part of the Hamiltonian as 
$H_0 = -t\sum_{m\sigma} (d^\dagger_{1m\sigma} d_{2m\sigma} + h.c.) = -2tT^x_\text{site}$, which connects through virtual hopping the different $T_\text{site}=1$ triplet states. This process stabilizes them with respect to the site singlets by an energy $\approx -4t^2/\Delta_a$, where $\Delta_a$ is the energy cost of creating the specific atomic multiplet involved in the virtual hopping. 
Moreover, as long as $J/U<1/3$, we expect that the ground state in the high-U limit will localize the electrons in each site, reducing the Hilbert space to states with $T_{\text{site}}^z=0$. 
Thus, the superexchange interaction will favor the 6 states with $T_{\text{site}}=1$ and $T_{\text{site}}^z=0$, for which we will have the energy splitting
given by the atomic gaps $\Delta_a$ of the local $N=2$ multiplets in Eq.~\eqref{eq:E2multiplets}.

At $J=0$ the six states are degenerate: they form the antisymmetric sextet of SU(4). At finite $J<U/3$ the spin-triplet, orbital-singlet
bond has lowest energy. This means that the dimer is ferromagnetic in spin and antiferro-orbital in agreement with the  Kugel--Khomskii superexchange picture of the quarter-filled two-orbital model~\cite{kugelJahnTellerEffectMagnetism1982,heldMicroscopicConditionsFavoring1998,petersOrbitalMagneticOrder2010}.

Table~\ref{tab:dimer} summarizes the values of the cluster observables in this limit.

\begin{table}[tb]
\caption{Cluster observables in the dimer limit $t/\Delta_a\to0$, for an
equal-weight mixture over its lower energy manifold. Entropies are measured in units of $\ln 2$ (bits).}
\label{tab:dimer}
\small
\begin{tabular}{lcc}
\hline\hline
 & $J=0$  & $0<J<U/3$  \\
\hline
$S(\rho_{12})$ & $\log_2 6$ & $\log_2 3$ \\
$\Imut$ & $\log_2(8/3)$ & $\log_2(16/3)$ \\
$\langle S^z_iS^z_j\rangle$ & $-1/12$ & $+1/12$ \\
$C^{ij}_{mm}$ & $-1/12$ & $-1/4$ \\
$C^{ij}_{mm'}$ & $+1/12$ & $+1/4$ \\
$\langle\tau^z_i\tau^z_j\rangle$ & $-1/12$ & $-1/4$ \\
\hline\hline
\end{tabular}
\end{table}

\subsection{Local observables: comparison with single-site DMFT}
\label{sec:cluster-local}

\begin{figure}[t]
  \centering
  \includegraphics[width=\columnwidth]{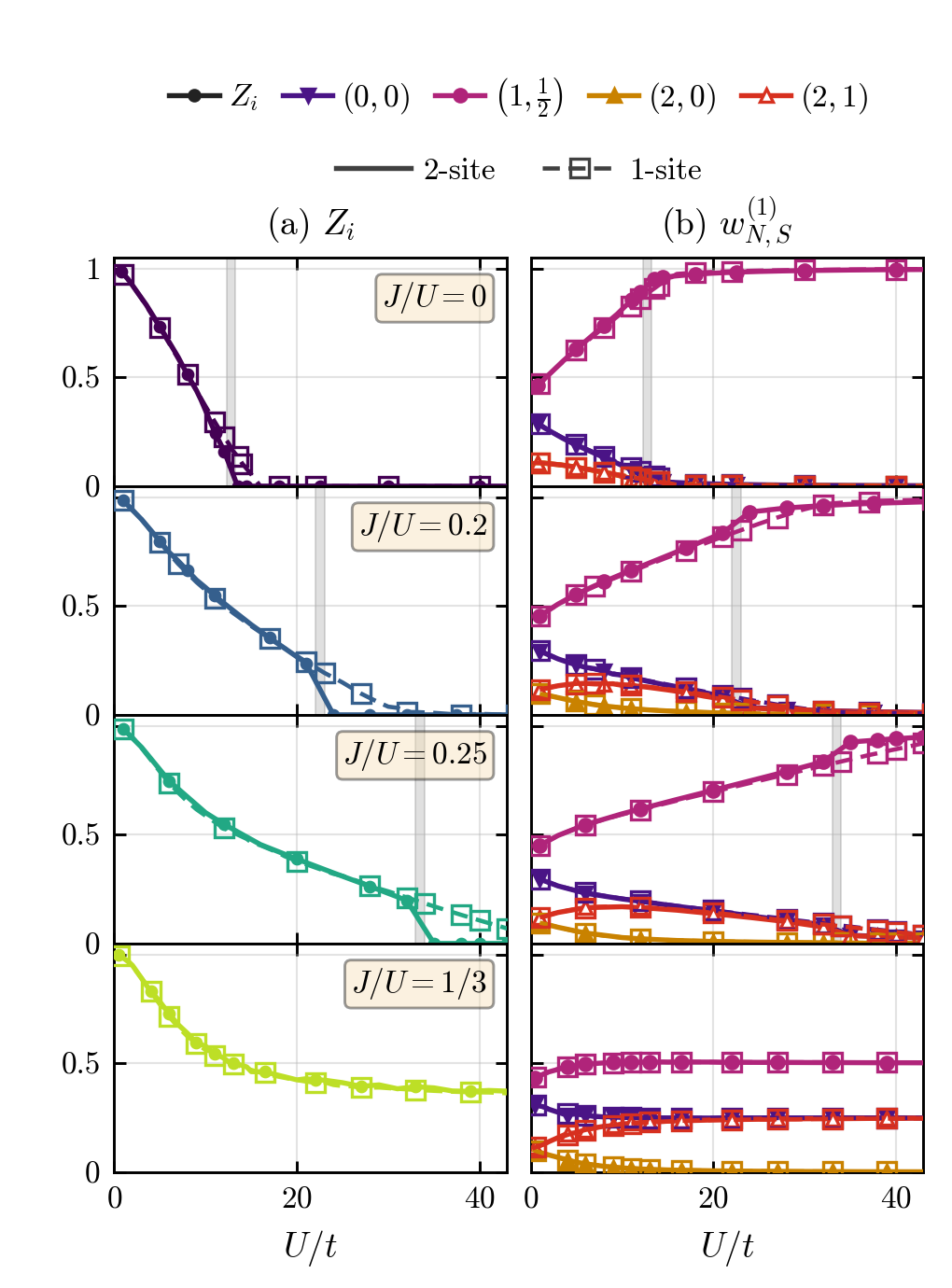}
  \caption{Quasiparticle weights $Z_i$ and multiplet weights $\wNS$ from the
  single-site and two-site cluster DMFT calculations at $\delta=0$: the local
  observables are nearly unchanged by the cluster.}
  \label{fig:cluster-local}
\end{figure}

Figure~\ref{fig:cluster-local} compares, at $\delta=0$, the on-site quasiparticle weight $Z_i$ and the multiplet weights of a cluster site, $w^{(i)}_{N,S}=\mathrm{Tr}(\rho_iP_{N,S})$, with the single-site results. We recall that $\rho_i$ is the reduced density matrix of cluster site $i$ (App.~\ref{app:rdm}). 

It is clear from the plots that the weights  of the multiplets of the CDMFT agree closely with DMFT over the whole range of $U$, and $Z_i$ agrees with the single-site $Z$ as long as the system is metallic in both methods.
The two-site CDMFT solution, however, displays a smaller critical $U$ for the Mott transition. This shift of the localization to smaller interaction is consistent with the effect of short-range correlations found in cluster studies of the single-band model~\cite{parkClusterDynamicalMean2008,balzerFirstorderMottTransition2009,meixnerMottTransitionPseudogap2024}. Besides this effect, we can conclude that the local physics of the quarter-filled two-orbital Hubbard model is robust to the inclusion of nearest-neighbor correlations.


\subsection{Intersite correlations}
\label{sec:correlators}

\begin{figure}[t]
  \centering
  \includegraphics[width=\columnwidth]{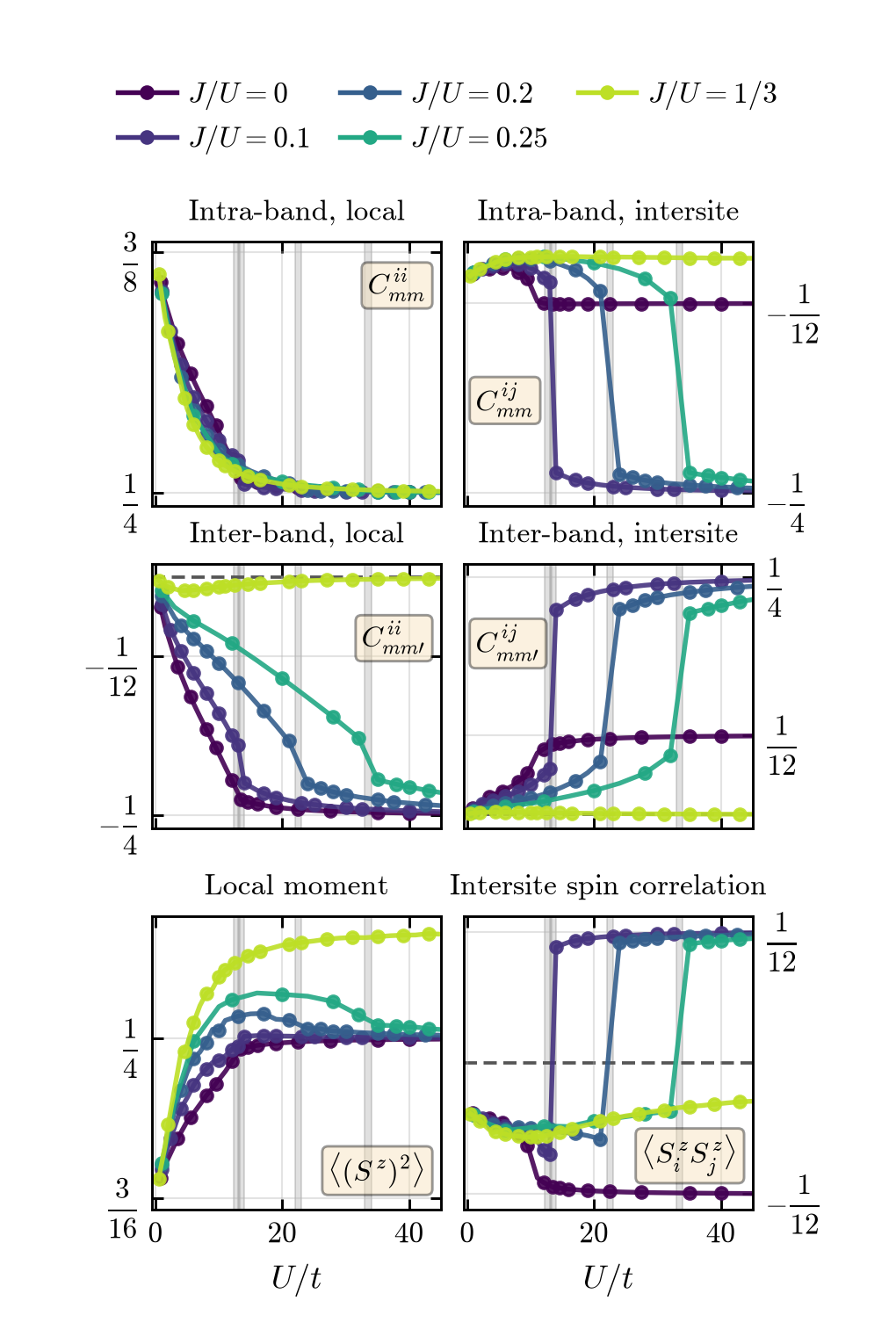}
  \caption{Local (left) and intersite (right) correlators of the two-site
  cluster at $\delta=0$ against $U/t$: intra-band (top) and inter-band (middle)
  charge correlators $C^{ij}_{mm'}$ (Eq.~\eqref{eq:Cij}), and the local moment
  $\langle(S^z_i)^2\rangle$ and intersite spin correlation
  $\langle S_i^z S_j^z\rangle$ (bottom). Dashed horizontal lines mark zero, while the
  labelled ticks are the limits of Table~\ref{tab:dimer}.}
  \label{fig:rho-observables-vs-U}
\end{figure}

Figure~\ref{fig:rho-observables-vs-U} shows the local and
intersite charge and spin correlators as functions of $U/t$
for different values of $\JU$.
The main distinction is between the metallic and localized
regimes: intersite correlations remain weak and depend little
on $J$ throughout the metal, whereas localization produces
a pronounced reorganization of the bond correlations.

The intersite charge correlators satisfy $C^{ij}_{mm}<0$ and $C^{ij}_{mm'}>0$ for $m\neq m'$ and all the values $\JU<1/3$ considered here.
This indicates antiferro-orbital correlations, which
are already present in the metal but become much stronger
upon localization reaching maximum values in the insulator.

The intersite spin correlation provides a sharper distinction
between the two regimes.
For finite $J$, $\langle S_i^z S_j^z\rangle$ remains weakly
antiferromagnetic throughout the metal and becomes
ferromagnetic at localization.
The resulting ferromagnetic, antiferro-orbital bond is
consistent with the Kugel--Khomskii superexchange discussed
in Sec.~\ref{sec:dimer}.

The case $J=0$ is different because of the SU(4) symmetry,
which implies equal spin and orbital correlations.
The low-energy dimer manifold is then the SU(4) sextet,
with antiferromagnetic correlations in both sectors.
At large $U$, the intersite correlators approach
$C^{ij}_{mm}\to-1/12$,
$C^{ij}_{mm'}\to+1/12$, and
$\langle S_i^z S_j^z\rangle\to-1/12$,
in agreement with Table~\ref{tab:dimer}.
Thus, although the local atomic state is singly occupied
for both $J=0$ and finite $J$, the Hund's coupling changes
the spin character of the insulating bond.

The local observables provide a complementary picture.
At intermediate $U$, finite $J$ favors the inter-band
triplet over the intra-band singlets in the doublon sector.
The associated triplet weight
[Fig.~\ref{fig:cluster-local}] enhances the local moment,
which can exceed the singly occupied value
$\langle(S_i^z)^2\rangle=1/4$.
As charge fluctuations are suppressed at larger $U$,
the local correlators approach
$C^{ii}_{mm}\to1/4$,
$C^{ii}_{mm'}\to-1/4$, and
$\langle(S_i^z)^2\rangle\to1/4$.
For $\JU<1/3$, these limiting values are independent of $J$.
Their similarity to the single-site results is consistent
with the close agreement of the local multiplet weights
shown in Fig.~\ref{fig:cluster-local}.
The strong $J$ dependence of the insulating bond therefore
coexists with essentially unchanged local observables.

\begin{figure}[t]
  \centering
  \includegraphics[width=\columnwidth]{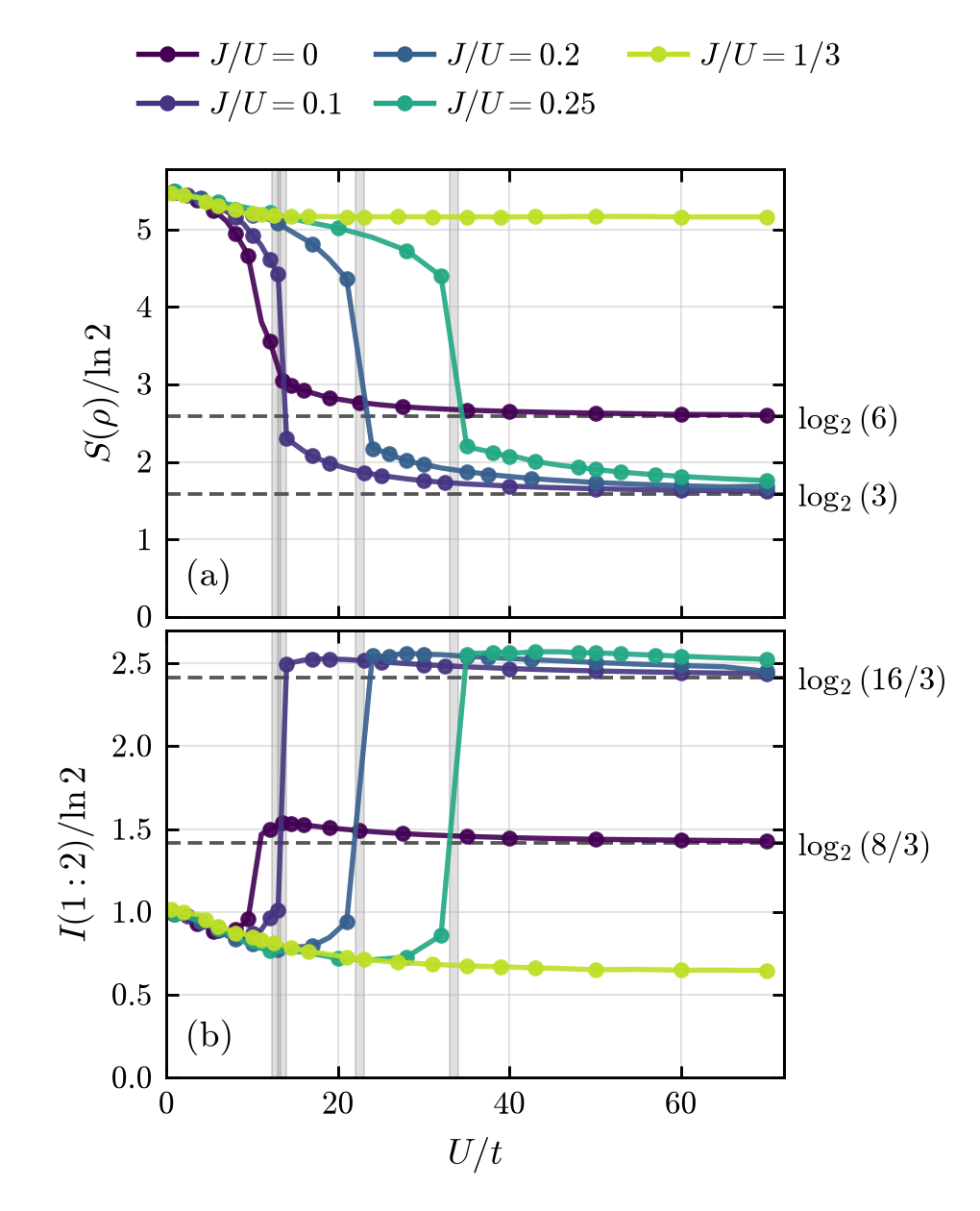}
  \caption{Top: von Neumann entropy $S(\rho_{12})/\ln2$ of the two-site cluster
  reduced density matrix, with the reference values $\log_2 6$ and $\log_2 3$.
  Bottom: intersite mutual information $\Imut/\ln 2$, with the reference values
  $\log_2(8/3)$ and $\log_2(16/3)$ (Table~\ref{tab:dimer}). Both in bits, against
  $U/t$, for the values of $\JU$ in the legend.}
  \label{fig:intersite-mutual-information}
\end{figure}

The bond reorganization is also visible in the cluster
entropy and intersite mutual information,
shown in Fig.~\ref{fig:intersite-mutual-information}.
At $U=0$, $\Imut/\ln 2\approx1$ bit, reflecting correlations
generated by hopping even in the noninteracting system.
The mutual information decreases slowly with increasing
$U$ within the metal, then rises abruptly at localization.

For finite $J$, the insulating mutual information reaches
a plateau slightly above
$\log_2(16/3)\approx2.415$ bits, while
$S(\rho_{12})/\ln 2$ approaches $\log_2 3$ from above.
These are the reference values for an equal-weight mixture
over the spin-triplet, orbital-singlet dimer manifold
(Table~\ref{tab:dimer}).
At $J=0$, the corresponding limits are instead
$\Imut/\ln 2\to\log_2(8/3)\approx1.415$ bits and
$S(\rho_{12})/\ln 2\to\log_2 6$,
as expected for the equal-weight SU(4) sextet.
The deviations from these ideal dimer values reflect
residual charge fluctuations and coupling to the bath.
Because the mutual information combines the site and
cluster entropies, its deviation depends on the balance
between both contributions.

At $\JU=1/3$, neither quantity displays the localization
jump found for $\JU<1/3$.
Instead, the cluster entropy remains large and both
quantities decrease slowly over the interaction range
studied.
This behavior accompanies the persistence of the metallic
solution at vanishing atomic charge gap.

In the metallic regime, we find no indication of the
Fermi-liquid breakdown and pseudogap reported by the NCA
study of Ref.~\cite{kitaSpatialFluctuationsSpin2009}
at $U=4t$.
For $\JU=0.2$--$0.25$, our $T=0$ solutions have
$Z_i\approx0.85$ [Fig.~\ref{fig:cluster-local}],
while the intersite spin and orbital correlations are
weak and nearly independent of $J$
[Fig.~\ref{fig:rho-observables-vs-U}].

The two-site calculation thus preserves the local Hund
signatures of the single-site metal while resolving
the correlations carried by the explicit bond.
Within the metal, these correlations remain weak and
nearly independent of $J$.
At localization, they change abruptly and approach the
isolated-dimer limits: an antiferromagnetic spin--orbital
bond at $J=0$, and a ferromagnetic, antiferro-orbital bond
at finite $J$.
The correlators and mutual information therefore identify
a substantial reorganization of the joint two-site state
that is only weakly reflected in the local observables.
These local observables are also unchanged by the 2-site calculation with respect to the single-site, since they are fully controlled by the value of the single site reduced density matrix.

\section{Conclusion}
\label{sec:conclusion}

In this work we have studied the quarter-filled two-band Hubbard--Kanamori model with full interaction including spin-flip and pair-hopping terms by means of zero-temperature single-site DMFT and two-site cellular DMFT.

Our main focus is on the non-trivial signatures of the Hund's coupling $J$ in this regime, where each site hosts only one electron. $J$ leaves the singly occupied atomic multiplet unsplit,  but it splits the $N=2$ manifold which plays the role of a virtual state. As a result, both the quasiparticle weight and the structure of local multiplets are affected by the Hund's exchange coupling and the system can be described as a weak Hund's metal.

In this work, in the perspective of calculations for materials, we discussed the robustness of these results with respect to a crystal field splitting breaking the degeneracy between the two orbitals and non-local correlations introduced in cluster DMFT.

A crystal-field splitting leads to an orbital polarization, which is expected to compete with the Hund's coupling, and makes the response orbital dependent. We find indeed that the orbital polarization opposes the Hund's driven suppression of coherence in the more occupied band while it reinforces the effect in the less occupied one. 
Very interestingly, we find that the signatures of the Hund's coupling in the evolution of the quasiparticle weight are quite robust as they persist to a splitting $\delta=2t$, in contrast with previous reports that show the disappearance of a two-stage Kondo screening already for $\delta=t$. Thus, fingerprints of the weak Hund's metal can exist even when the spin-orbital separation is absent, suggesting that the Hund's driven correlation can be subtler than expected. 

We have also studied the effects of non-local correlation by considering the model within two-site cellular DMFT.
We find that the local multiplets and the quasiparticle weights are weakly affected by non-local correlations. As a matter of fact the non-local correlations are essentially irrelevant in the metallic state and their main effect is to reduce the critical $U$ for the Mott transition. In the insulating state the non-local effects select  a ferromagnetic, antiferro-orbital bond of Kugel--Khomskii type. The latter is indeed the ground state of an isolated dimer. All the intersite correlators and the mutual information confirm this picture as they approach the values for an isolated cluster.

Our results show that the Hund's exchange effects at quarter filling are more subtle than expected and the scenario emergent from DMFT turns out to be remarkably robust with respect both to a crystal-field splitting and to non-local correlations. The somewhat surprising robustness of the weak Hund's metal suggests that this state is likely to be observed also in real materials which are not described by the simple two-orbital model in DMFT approximation, calling for experimental investigations of this scenario.

\begin{acknowledgments}
We acknowledge useful discussions with G. Bellomia, E. Linnèr, E. Zavatti. High performance calculations were carried out thanks to resources provided by CINECA HPC via Convenzione SISSA 2025 and 2026.
\end{acknowledgments}

\appendix

\section{Reduced density matrices and projectors}
\label{app:rdm}

We expand the ground state of the impurity model in Slater determinants,
\begin{equation}
    |\Psi\rangle
    =
    \sum_{\ell,\alpha}
    \lambda_{\ell\alpha}\,
    |\ell,\alpha\rangle ,
\end{equation}
where $\ell$ ($\alpha$) labels the occupations of the impurity (bath) spin orbitals. In $|\ell,\alpha\rangle$ the creation operators act on the vacuum in the following order: all spin-up orbitals precede all spin-down ones, and within each spin block the impurity orbitals precede the bath orbitals,
\begin{equation}
    [\,\mathrm{imp}{\uparrow}\,]\,
    [\,\mathrm{bath}{\uparrow}\,]\,
    [\,\mathrm{imp}{\downarrow}\,]\,
    [\,\mathrm{bath}{\downarrow}\,] .
    \label{eq:ordering}
\end{equation}
In single-site DMFT the impurity block is ordered simply by band, $m=1,2$. In the two-site cluster it is ordered according to the convention: $1\equiv(\text{site } 1,\text{band } 1)$, $2\equiv(\text{site } 2,\text{band } 1)$, $3\equiv(\text{site } 1,\text{band } 2)$, and $4\equiv(\text{site } 2,\text{band } 2)$.

The impurity reduced density matrix is $\hat{\rho}_{\mathrm{imp}}=\operatorname{Tr}_{\mathrm{bath}}|\Psi\rangle\langle\Psi|$. Before the bath is traced out, its operators must be moved past the impurity operators with which they are interleaved in Eq.~\eqref{eq:ordering}. Moving them to the left multiplies $|\ell,\alpha\rangle$ by $(-1)^{\varphi(\ell,\alpha)}$, with
\begin{equation}
    \varphi(\ell,\alpha)
    =
    n_{b\uparrow}\,n_{I\uparrow}
    +
    n_{b\downarrow}\,n_{I} ,
\end{equation}
where $n_{b\sigma}$ ($n_{I\sigma}$) is the number of bath (impurity) electrons with spin $\sigma$ and $n_I=n_{I\uparrow}+n_{I\downarrow}$. The matrix elements are
\begin{equation}
    \rho^{\mathrm{imp}}_{\ell'\ell}
    =
    \sum_{\alpha}
    (-1)^{\varphi(\ell,\alpha)+\varphi(\ell',\alpha)}\,
    \lambda_{\ell'\alpha}\,
    \lambda^{*}_{\ell\alpha} .
\end{equation}
For the two-site cluster, $\hat{\rho}_{\mathrm{imp}}$ is the cluster density matrix $\rho_{12}$ of Sec.~\ref{sec:cluster}.

The density matrix of site 1 is $\hat{\rho}_1=\operatorname{Tr}_2\hat{\rho}_{12}$. We split the cluster occupations as $\ell=(\ell_1,\ell_2)$, with $\ell_i$ those of site $i$. Within each spin block of Eq.~\eqref{eq:ordering} the orbitals of the two sites alternate. Moving the site-1 operators of $|\ell_1\ell_2\rangle$ to the left of the site-2 ones therefore multiplies it by $(-1)^{\varphi_{12}(\ell_1,\ell_2)}$, with
\begin{equation}
    \varphi_{12}
    =
    \sum_{\sigma}
    n_{3\sigma}\,m_{2\sigma}
    +
    \left(n_{1\downarrow}+n_{3\downarrow}\right)
    \left(m_{2\uparrow}+m_{4\uparrow}\right) ,
    \label{eq:phi12}
\end{equation}
where $n_{k\sigma}$ are the occupations of the site-1 orbitals ($k=1,3$) in $\ell_1$ and $m_{k\sigma}$ those of the site-2 orbitals ($k=2,4$) in $\ell_2$. The first term counts the exchanges of orbital $3\sigma$ with orbital $2\sigma$ within each spin block, and the second those of the site-1 spin-down operators with the site-2 spin-up ones. The matrix elements are
\begin{equation}
    \rho^{(1)}_{\ell'_1\ell_1}
    =
    \sum_{\ell_2}
    (-1)^{\varphi_{12}(\ell_1,\ell_2)+\varphi_{12}(\ell'_1,\ell_2)}\,
    \rho^{(12)}_{(\ell'_1\ell_2),(\ell_1\ell_2)} ,
    \label{eq:rho1}
\end{equation}
and $\rho_2$ follows by summing over $\ell_1$ with the same sign. 

The multiplet weights are the expectation values of projectors onto the local atomic multiplets, which we build with L\"owdin's projection operators~\cite{lowdinQuantumTheoryManyParticle1955,lowdinAngularMomentumWavefunctions1964}. For commuting operators $O_i$ with known spectra, the projector onto their joint eigenspace with eigenvalues $q_i$ is
\begin{equation}
    P_{\{q_i\}}
    =
    \prod_{i}
    \prod_{q'_i\neq q_i}
    \frac{O_i-q'_i}{q_i-q'_i} ,
\end{equation}
where $q'_i$ runs over the other eigenvalues of $O_i$. For $P_{N,S}$ we take $O=\{N,\bm{S}^2\}$ on the 16-dimensional Fock space of one site, with $N\in\{0,\ldots,4\}$ and $\bm{S}^2$ eigenvalues $S(S+1)\in\{0,3/4,2\}$. The six nonempty sectors $(N,S)=(0,0)$, $(1,\tfrac12)$, $(2,0)$, $(2,1)$, $(3,\tfrac12)$, and $(4,0)$ have dimensions $1,4,3,3,4,1$. The weights are
\begin{equation}
    \wNS
    =
    \mathrm{Tr}\left(\rho\, P_{N,S}\right) ,
\end{equation}
with $\rho=\rho_{\mathrm{imp}}$ in single-site DMFT and $\rho=\rho_i$ for cluster site $i$, and they satisfy $\sum_{N,S}\wNS=1$.



%% file: paper.bbl
\begin{thebibliography}{44}%
\makeatletter
\providecommand \@ifxundefined [1]{%
 \@ifx{#1\undefined}
}%
\providecommand \@ifnum [1]{%
 \ifnum #1\expandafter \@firstoftwo
 \else \expandafter \@secondoftwo
 \fi
}%
\providecommand \@ifx [1]{%
 \ifx #1\expandafter \@firstoftwo
 \else \expandafter \@secondoftwo
 \fi
}%
\providecommand \natexlab [1]{#1}%
\providecommand \enquote  [1]{``#1''}%
\providecommand \bibnamefont  [1]{#1}%
\providecommand \bibfnamefont [1]{#1}%
\providecommand \citenamefont [1]{#1}%
\providecommand \href@noop [0]{\@secondoftwo}%
\providecommand \href [0]{\begingroup \@sanitize@url \@href}%
\providecommand \@href[1]{\@@startlink{#1}\@@href}%
\providecommand \@@href[1]{\endgroup#1\@@endlink}%
\providecommand \@sanitize@url [0]{\catcode `\\12\catcode `\$12\catcode
  `\&12\catcode `\#12\catcode `\^12\catcode `\_12\catcode `\%12\relax}%
\providecommand \@@startlink[1]{}%
\providecommand \@@endlink[0]{}%
\providecommand \url  [0]{\begingroup\@sanitize@url \@url }%
\providecommand \@url [1]{\endgroup\@href {#1}{\urlprefix }}%
\providecommand \urlprefix  [0]{URL }%
\providecommand \Eprint [0]{\href }%
\providecommand \doibase [0]{https://doi.org/}%
\providecommand \selectlanguage [0]{\@gobble}%
\providecommand \bibinfo  [0]{\@secondoftwo}%
\providecommand \bibfield  [0]{\@secondoftwo}%
\providecommand \translation [1]{[#1]}%
\providecommand \BibitemOpen [0]{}%
\providecommand \bibitemStop [0]{}%
\providecommand \bibitemNoStop [0]{.\EOS\space}%
\providecommand \EOS [0]{\spacefactor3000\relax}%
\providecommand \BibitemShut  [1]{\csname bibitem#1\endcsname}%
\let\auto@bib@innerbib\@empty
\bibitem [{\citenamefont {Georges}\ \emph {et~al.}(2013)\citenamefont
  {Georges}, \citenamefont {Medici},\ and\ \citenamefont
  {Mravlje}}]{georgesStrongCorrelationsHunds2013}%
  \BibitemOpen
  \bibfield  {author} {\bibinfo {author} {\bibfnamefont {A.}~\bibnamefont
  {Georges}}, \bibinfo {author} {\bibfnamefont {L.~D.}\ \bibnamefont
  {Medici}},\ and\ \bibinfo {author} {\bibfnamefont {J.}~\bibnamefont
  {Mravlje}},\ }\bibfield  {title} {\bibinfo {title} {Strong {{Correlations}}
  from {{Hund}}'s {{Coupling}}},\ }\href
  {https://doi.org/10.1146/annurev-conmatphys-020911-125045} {\bibfield
  {journal} {\bibinfo  {journal} {Annual Review of Condensed Matter Physics}\
  }\textbf {\bibinfo {volume} {4}},\ \bibinfo {pages} {137} (\bibinfo {year}
  {2013})}\BibitemShut {NoStop}%
\bibitem [{\citenamefont {Haule}\ and\ \citenamefont
  {Kotliar}(2009)}]{hauleCoherenceIncoherenceCrossover2009}%
  \BibitemOpen
  \bibfield  {author} {\bibinfo {author} {\bibfnamefont {K.}~\bibnamefont
  {Haule}}\ and\ \bibinfo {author} {\bibfnamefont {G.}~\bibnamefont
  {Kotliar}},\ }\bibfield  {title} {\bibinfo {title} {Coherence--incoherence
  crossover in the normal state of iron oxypnictides and importance of
  {{Hund}}'s rule coupling},\ }\href
  {https://doi.org/10.1088/1367-2630/11/2/025021} {\bibfield  {journal}
  {\bibinfo  {journal} {New Journal of Physics}\ }\textbf {\bibinfo {volume}
  {11}},\ \bibinfo {pages} {025021} (\bibinfo {year} {2009})}\BibitemShut
  {NoStop}%
\bibitem [{\citenamefont {Werner}\ \emph {et~al.}(2008)\citenamefont {Werner},
  \citenamefont {Gull}, \citenamefont {Troyer},\ and\ \citenamefont
  {Millis}}]{wernerSpinFreezingTransition2008}%
  \BibitemOpen
  \bibfield  {author} {\bibinfo {author} {\bibfnamefont {P.}~\bibnamefont
  {Werner}}, \bibinfo {author} {\bibfnamefont {E.}~\bibnamefont {Gull}},
  \bibinfo {author} {\bibfnamefont {M.}~\bibnamefont {Troyer}},\ and\ \bibinfo
  {author} {\bibfnamefont {A.~J.}\ \bibnamefont {Millis}},\ }\bibfield  {title}
  {\bibinfo {title} {Spin freezing transition and non-{{Fermi}}-liquid
  self-energy in a three-orbital model},\ }\href
  {https://doi.org/10.1103/PhysRevLett.101.166405} {\bibfield  {journal}
  {\bibinfo  {journal} {Physical Review Letters}\ }\textbf {\bibinfo {volume}
  {101}},\ \bibinfo {pages} {166405} (\bibinfo {year} {2008})}\BibitemShut
  {NoStop}%
\bibitem [{\citenamefont {Georges}\ and\ \citenamefont
  {Kotliar}(2024)}]{georgesHundmetalPathStrong2024}%
  \BibitemOpen
  \bibfield  {author} {\bibinfo {author} {\bibfnamefont {A.}~\bibnamefont
  {Georges}}\ and\ \bibinfo {author} {\bibfnamefont {G.}~\bibnamefont
  {Kotliar}},\ }\bibfield  {title} {\bibinfo {title} {The {{Hund-metal}} path
  to strong electronic correlations},\ }\href
  {https://doi.org/10.1063/pt.wqrz.qpjx} {\bibfield  {journal} {\bibinfo
  {journal} {Physics Today}\ }\textbf {\bibinfo {volume} {77}},\ \bibinfo
  {pages} {46} (\bibinfo {year} {2024})}\BibitemShut {NoStop}%
\bibitem [{\citenamefont {Capone}\ and\ \citenamefont
  {Fanfarillo}(2026)}]{Capone2026}%
  \BibitemOpen
  \bibfield  {author} {\bibinfo {author} {\bibfnamefont {M.}~\bibnamefont
  {Capone}}\ and\ \bibinfo {author} {\bibfnamefont {L.}~\bibnamefont
  {Fanfarillo}},\ }\bibfield  {title} {\bibinfo {title} {Hund's metals: the
  interplay of correlations and emergent orders},\ }\href
  {https://doi.org/10.1080/00107514.2026.2647572} {\bibfield  {journal}
  {\bibinfo  {journal} {Contemporary Physics}\ }\textbf {\bibinfo {volume}
  {67}},\ \bibinfo {pages} {93} (\bibinfo {year} {2026})},\ \Eprint
  {https://arxiv.org/abs/https://doi.org/10.1080/00107514.2026.2647572}
  {https://doi.org/10.1080/00107514.2026.2647572} \BibitemShut {NoStop}%
\bibitem [{\citenamefont {De'~Medici}\ \emph {et~al.}(2011)\citenamefont
  {De'~Medici}, \citenamefont {Mravlje},\ and\ \citenamefont
  {Georges}}]{demediciJanusFacedInfluenceHunds2011}%
  \BibitemOpen
  \bibfield  {author} {\bibinfo {author} {\bibfnamefont {L.}~\bibnamefont
  {De'~Medici}}, \bibinfo {author} {\bibfnamefont {J.}~\bibnamefont
  {Mravlje}},\ and\ \bibinfo {author} {\bibfnamefont {A.}~\bibnamefont
  {Georges}},\ }\bibfield  {title} {\bibinfo {title} {Janus-{{Faced Influence}}
  of {{Hund}}'s {{Rule Coupling}} in {{Strongly Correlated Materials}}},\
  }\href {https://doi.org/10.1103/PhysRevLett.107.256401} {\bibfield  {journal}
  {\bibinfo  {journal} {Physical Review Letters}\ }\textbf {\bibinfo {volume}
  {107}},\ \bibinfo {pages} {256401} (\bibinfo {year} {2011})}\BibitemShut
  {NoStop}%
\bibitem [{\citenamefont {{de' Medici}}(2011)}]{demediciHundsCouplingIts2011}%
  \BibitemOpen
  \bibfield  {author} {\bibinfo {author} {\bibfnamefont {L.}~\bibnamefont {{de'
  Medici}}},\ }\bibfield  {title} {\bibinfo {title} {Hund's coupling and its
  key role in tuning multiorbital correlations},\ }\href
  {https://doi.org/10.1103/PhysRevB.83.205112} {\bibfield  {journal} {\bibinfo
  {journal} {Physical Review B}\ }\textbf {\bibinfo {volume} {83}},\ \bibinfo
  {pages} {205112} (\bibinfo {year} {2011})}\BibitemShut {NoStop}%
\bibitem [{\citenamefont {Stadler}\ \emph {et~al.}(2015)\citenamefont
  {Stadler}, \citenamefont {Yin}, \citenamefont {{von Delft}}, \citenamefont
  {Kotliar},\ and\ \citenamefont
  {Weichselbaum}}]{stadlerDynamicalMeanFieldTheory2015}%
  \BibitemOpen
  \bibfield  {author} {\bibinfo {author} {\bibfnamefont {K.~M.}\ \bibnamefont
  {Stadler}}, \bibinfo {author} {\bibfnamefont {Z.~P.}\ \bibnamefont {Yin}},
  \bibinfo {author} {\bibfnamefont {J.}~\bibnamefont {{von Delft}}}, \bibinfo
  {author} {\bibfnamefont {G.}~\bibnamefont {Kotliar}},\ and\ \bibinfo {author}
  {\bibfnamefont {A.}~\bibnamefont {Weichselbaum}},\ }\bibfield  {title}
  {\bibinfo {title} {Dynamical {{Mean-Field Theory Plus Numerical
  Renormalization-Group Study}} of {{Spin-Orbital Separation}} in a
  {{Three-Band Hund Metal}}},\ }\href
  {https://doi.org/10.1103/PhysRevLett.115.136401} {\bibfield  {journal}
  {\bibinfo  {journal} {Physical Review Letters}\ }\textbf {\bibinfo {volume}
  {115}},\ \bibinfo {pages} {136401} (\bibinfo {year} {2015})}\BibitemShut
  {NoStop}%
\bibitem [{\citenamefont {Ryee}\ \emph {et~al.}(2021)\citenamefont {Ryee},
  \citenamefont {Han},\ and\ \citenamefont
  {Choi}}]{ryeeHundPhysicsLandscape2021}%
  \BibitemOpen
  \bibfield  {author} {\bibinfo {author} {\bibfnamefont {S.}~\bibnamefont
  {Ryee}}, \bibinfo {author} {\bibfnamefont {M.~J.}\ \bibnamefont {Han}},\ and\
  \bibinfo {author} {\bibfnamefont {S.}~\bibnamefont {Choi}},\ }\bibfield
  {title} {\bibinfo {title} {Hund {{Physics Landscape}} of {{Two-Orbital
  Systems}}},\ }\href {https://doi.org/10.1103/PhysRevLett.126.206401}
  {\bibfield  {journal} {\bibinfo  {journal} {Physical Review Letters}\
  }\textbf {\bibinfo {volume} {126}},\ \bibinfo {pages} {206401} (\bibinfo
  {year} {2021})}\BibitemShut {NoStop}%
\bibitem [{\citenamefont {Kugler}\ \emph {et~al.}(2024)\citenamefont {Kugler},
  \citenamefont {Kang},\ and\ \citenamefont
  {Kotliar}}]{kuglerLowenergyPerspectiveTwoorbital2024}%
  \BibitemOpen
  \bibfield  {author} {\bibinfo {author} {\bibfnamefont {F.~B.}\ \bibnamefont
  {Kugler}}, \bibinfo {author} {\bibfnamefont {C.-J.}\ \bibnamefont {Kang}},\
  and\ \bibinfo {author} {\bibfnamefont {G.}~\bibnamefont {Kotliar}},\
  }\bibfield  {title} {\bibinfo {title} {Low-energy perspective on two-orbital
  {{Hund}} metals and the case of {{LaNiO2}}},\ }\href
  {https://doi.org/10.1103/PhysRevB.110.155101} {\bibfield  {journal} {\bibinfo
   {journal} {Physical Review B}\ }\textbf {\bibinfo {volume} {110}},\ \bibinfo
  {pages} {155101} (\bibinfo {year} {2024})},\ \Eprint
  {https://arxiv.org/abs/2312.11457} {arXiv:2312.11457 [cond-mat]} \BibitemShut
  {NoStop}%
\bibitem [{\citenamefont {Koga}\ \emph {et~al.}(2005)\citenamefont {Koga},
  \citenamefont {Inaba},\ and\ \citenamefont
  {Kawakami}}]{kogaMottTransitionsTwoOrbital2005}%
  \BibitemOpen
  \bibfield  {author} {\bibinfo {author} {\bibfnamefont {A.}~\bibnamefont
  {Koga}}, \bibinfo {author} {\bibfnamefont {K.}~\bibnamefont {Inaba}},\ and\
  \bibinfo {author} {\bibfnamefont {N.}~\bibnamefont {Kawakami}},\ }\bibfield
  {title} {\bibinfo {title} {Mott {{Transitions}} in {{Two-Orbital Hubbard
  Systems}}},\ }\href {https://doi.org/10.1143/PTPS.160.253} {\bibfield
  {journal} {\bibinfo  {journal} {Progress of Theoretical Physics Supplement}\
  }\textbf {\bibinfo {volume} {160}},\ \bibinfo {pages} {253} (\bibinfo {year}
  {2005})}\BibitemShut {NoStop}%
\bibitem [{\citenamefont {Inaba}\ and\ \citenamefont
  {Koga}(2007)}]{inabaMetalinsulatorTransitionTwoorbital2007}%
  \BibitemOpen
  \bibfield  {author} {\bibinfo {author} {\bibfnamefont {K.}~\bibnamefont
  {Inaba}}\ and\ \bibinfo {author} {\bibfnamefont {A.}~\bibnamefont {Koga}},\
  }\bibfield  {title} {\bibinfo {title} {Metal-insulator transition in the
  two-orbital {{Hubbard}} model at fractional band fillings: {{Self-energy}}
  functional approach},\ }\href {https://doi.org/10.1143/JPSJ.76.094712}
  {\bibfield  {journal} {\bibinfo  {journal} {Journal of the Physical Society
  of Japan}\ }\textbf {\bibinfo {volume} {76}},\ \bibinfo {pages} {094712}
  (\bibinfo {year} {2007})},\ \Eprint {https://arxiv.org/abs/0706.3948}
  {arXiv:0706.3948 [cond-mat]} \BibitemShut {NoStop}%
\bibitem [{\citenamefont {Held}\ and\ \citenamefont
  {Vollhardt}(1998)}]{heldMicroscopicConditionsFavoring1998}%
  \BibitemOpen
  \bibfield  {author} {\bibinfo {author} {\bibfnamefont {K.}~\bibnamefont
  {Held}}\ and\ \bibinfo {author} {\bibfnamefont {D.}~\bibnamefont
  {Vollhardt}},\ }\bibfield  {title} {\bibinfo {title} {Microscopic conditions
  favoring itinerant ferromagnetism: {{Hund}}'s rule coupling and orbital
  degeneracy},\ }\href {https://doi.org/10.1007/s100510050468} {\bibfield
  {journal} {\bibinfo  {journal} {The European Physical Journal B}\ }\textbf
  {\bibinfo {volume} {5}},\ \bibinfo {pages} {473} (\bibinfo {year}
  {1998})}\BibitemShut {NoStop}%
\bibitem [{\citenamefont {Peters}\ and\ \citenamefont
  {Pruschke}(2010)}]{petersOrbitalMagneticOrder2010}%
  \BibitemOpen
  \bibfield  {author} {\bibinfo {author} {\bibfnamefont {R.}~\bibnamefont
  {Peters}}\ and\ \bibinfo {author} {\bibfnamefont {T.}~\bibnamefont
  {Pruschke}},\ }\bibfield  {title} {\bibinfo {title} {Orbital and magnetic
  order in the two-orbital {{Hubbard}} model},\ }\href
  {https://doi.org/10.1103/PhysRevB.81.035112} {\bibfield  {journal} {\bibinfo
  {journal} {Physical Review B}\ }\textbf {\bibinfo {volume} {81}},\ \bibinfo
  {pages} {035112} (\bibinfo {year} {2010})}\BibitemShut {NoStop}%
\bibitem [{\citenamefont {Isidori}\ \emph {et~al.}(2019)\citenamefont
  {Isidori}, \citenamefont {Berovi{\'c}}, \citenamefont {Fanfarillo},
  \citenamefont {De'~Medici}, \citenamefont {Fabrizio},\ and\ \citenamefont
  {Capone}}]{isidoriChargeDisproportionationMixed2019}%
  \BibitemOpen
  \bibfield  {author} {\bibinfo {author} {\bibfnamefont {A.}~\bibnamefont
  {Isidori}}, \bibinfo {author} {\bibfnamefont {M.}~\bibnamefont
  {Berovi{\'c}}}, \bibinfo {author} {\bibfnamefont {L.}~\bibnamefont
  {Fanfarillo}}, \bibinfo {author} {\bibfnamefont {L.}~\bibnamefont
  {De'~Medici}}, \bibinfo {author} {\bibfnamefont {M.}~\bibnamefont
  {Fabrizio}},\ and\ \bibinfo {author} {\bibfnamefont {M.}~\bibnamefont
  {Capone}},\ }\bibfield  {title} {\bibinfo {title} {Charge
  {{Disproportionation}}, {{Mixed Valence}}, and {{Janus Effect}} in
  {{Multiorbital Systems}}: {{A Tale}} of {{Two Insulators}}},\ }\href
  {https://doi.org/10.1103/PhysRevLett.122.186401} {\bibfield  {journal}
  {\bibinfo  {journal} {Physical Review Letters}\ }\textbf {\bibinfo {volume}
  {122}},\ \bibinfo {pages} {186401} (\bibinfo {year} {2019})}\BibitemShut
  {NoStop}%
\bibitem [{\citenamefont {Kitatani}\ \emph {et~al.}(2020)\citenamefont
  {Kitatani}, \citenamefont {Si}, \citenamefont {Janson}, \citenamefont
  {Arita}, \citenamefont {Zhong},\ and\ \citenamefont
  {Held}}]{kitataniNickelateSuperconductorsRenaissance2020}%
  \BibitemOpen
  \bibfield  {author} {\bibinfo {author} {\bibfnamefont {M.}~\bibnamefont
  {Kitatani}}, \bibinfo {author} {\bibfnamefont {L.}~\bibnamefont {Si}},
  \bibinfo {author} {\bibfnamefont {O.}~\bibnamefont {Janson}}, \bibinfo
  {author} {\bibfnamefont {R.}~\bibnamefont {Arita}}, \bibinfo {author}
  {\bibfnamefont {Z.}~\bibnamefont {Zhong}},\ and\ \bibinfo {author}
  {\bibfnamefont {K.}~\bibnamefont {Held}},\ }\bibfield  {title} {\bibinfo
  {title} {Nickelate superconductors---a renaissance of the one-band
  {{Hubbard}} model},\ }\href {https://doi.org/10.1038/s41535-020-00260-y}
  {\bibfield  {journal} {\bibinfo  {journal} {npj Quantum Materials}\ }\textbf
  {\bibinfo {volume} {5}},\ \bibinfo {pages} {1} (\bibinfo {year}
  {2020})}\BibitemShut {NoStop}%
\bibitem [{\citenamefont {Si}\ \emph {et~al.}(2024)\citenamefont {Si},
  \citenamefont {Jacob}, \citenamefont {Wu}, \citenamefont {Hausoel},
  \citenamefont {Krsnik}, \citenamefont {Worm}, \citenamefont {Di~Cataldo},
  \citenamefont {Janson},\ and\ \citenamefont
  {Held}}]{siClosingPossibleScenarios2024}%
  \BibitemOpen
  \bibfield  {author} {\bibinfo {author} {\bibfnamefont {L.}~\bibnamefont
  {Si}}, \bibinfo {author} {\bibfnamefont {E.}~\bibnamefont {Jacob}}, \bibinfo
  {author} {\bibfnamefont {W.}~\bibnamefont {Wu}}, \bibinfo {author}
  {\bibfnamefont {A.}~\bibnamefont {Hausoel}}, \bibinfo {author} {\bibfnamefont
  {J.}~\bibnamefont {Krsnik}}, \bibinfo {author} {\bibfnamefont
  {P.}~\bibnamefont {Worm}}, \bibinfo {author} {\bibfnamefont {S.}~\bibnamefont
  {Di~Cataldo}}, \bibinfo {author} {\bibfnamefont {O.}~\bibnamefont {Janson}},\
  and\ \bibinfo {author} {\bibfnamefont {K.}~\bibnamefont {Held}},\ }\bibfield
  {title} {\bibinfo {title} {Closing in on possible scenarios for
  infinite-layer nickelates: {{Comparison}} of dynamical mean-field theory with
  angular-resolved photoemission spectroscopy},\ }\href
  {https://doi.org/10.1103/PhysRevResearch.6.043104} {\bibfield  {journal}
  {\bibinfo  {journal} {Physical Review Research}\ }\textbf {\bibinfo {volume}
  {6}},\ \bibinfo {pages} {043104} (\bibinfo {year} {2024})}\BibitemShut
  {NoStop}%
\bibitem [{\citenamefont {Wang}\ \emph {et~al.}(2020)\citenamefont {Wang},
  \citenamefont {Kang}, \citenamefont {Miao},\ and\ \citenamefont
  {Kotliar}}]{wangHundsMetalPhysics2020}%
  \BibitemOpen
  \bibfield  {author} {\bibinfo {author} {\bibfnamefont {Y.}~\bibnamefont
  {Wang}}, \bibinfo {author} {\bibfnamefont {C.-J.}\ \bibnamefont {Kang}},
  \bibinfo {author} {\bibfnamefont {H.}~\bibnamefont {Miao}},\ and\ \bibinfo
  {author} {\bibfnamefont {G.}~\bibnamefont {Kotliar}},\ }\bibfield  {title}
  {\bibinfo {title} {Hund's metal physics: {{From SrNiO}} 2 to {{LaNiO}} 2},\
  }\href {https://doi.org/10.1103/PhysRevB.102.161118} {\bibfield  {journal}
  {\bibinfo  {journal} {Physical Review B}\ }\textbf {\bibinfo {volume}
  {102}},\ \bibinfo {pages} {161118} (\bibinfo {year} {2020})}\BibitemShut
  {NoStop}%
\bibitem [{\citenamefont {Kang}\ and\ \citenamefont
  {Kotliar}(2021)}]{kangOpticalPropertiesInfiniteLayer2021}%
  \BibitemOpen
  \bibfield  {author} {\bibinfo {author} {\bibfnamefont {C.-J.}\ \bibnamefont
  {Kang}}\ and\ \bibinfo {author} {\bibfnamefont {G.}~\bibnamefont {Kotliar}},\
  }\bibfield  {title} {\bibinfo {title} {Optical {{Properties}} of the
  {{Infinite-Layer La}} 1 - x {{Sr}} x {{NiO}} 2 and {{Hidden Hund}}'s
  {{Physics}}},\ }\href {https://doi.org/10.1103/PhysRevLett.126.127401}
  {\bibfield  {journal} {\bibinfo  {journal} {Physical Review Letters}\
  }\textbf {\bibinfo {volume} {126}},\ \bibinfo {pages} {127401} (\bibinfo
  {year} {2021})}\BibitemShut {NoStop}%
\bibitem [{\citenamefont {Kang}\ \emph {et~al.}(2023)\citenamefont {Kang},
  \citenamefont {Melnick}, \citenamefont {Semon}, \citenamefont {Ryee},
  \citenamefont {Han}, \citenamefont {Kotliar},\ and\ \citenamefont
  {Choi}}]{kangInfinitelayerNickelatesNieg2023}%
  \BibitemOpen
  \bibfield  {author} {\bibinfo {author} {\bibfnamefont {B.}~\bibnamefont
  {Kang}}, \bibinfo {author} {\bibfnamefont {C.}~\bibnamefont {Melnick}},
  \bibinfo {author} {\bibfnamefont {P.}~\bibnamefont {Semon}}, \bibinfo
  {author} {\bibfnamefont {S.}~\bibnamefont {Ryee}}, \bibinfo {author}
  {\bibfnamefont {M.~J.}\ \bibnamefont {Han}}, \bibinfo {author} {\bibfnamefont
  {G.}~\bibnamefont {Kotliar}},\ and\ \bibinfo {author} {\bibfnamefont
  {S.}~\bibnamefont {Choi}},\ }\bibfield  {title} {\bibinfo {title}
  {Infinite-layer nickelates as {{Ni-eg Hund}}'s metals},\ }\href
  {https://doi.org/10.1038/s41535-023-00568-5} {\bibfield  {journal} {\bibinfo
  {journal} {npj Quantum Materials}\ }\textbf {\bibinfo {volume} {8}},\
  \bibinfo {pages} {35} (\bibinfo {year} {2023})}\BibitemShut {NoStop}%
\bibitem [{\citenamefont
  {Lechermann}(2020)}]{lechermannMultiorbitalProcessesRule2020}%
  \BibitemOpen
  \bibfield  {author} {\bibinfo {author} {\bibfnamefont {F.}~\bibnamefont
  {Lechermann}},\ }\bibfield  {title} {\bibinfo {title} {Multiorbital
  {{Processes Rule}} the {{Nd}} 1 - x {{Sr}} x {{NiO}} 2 {{Normal State}}},\
  }\href {https://doi.org/10.1103/PhysRevX.10.041002} {\bibfield  {journal}
  {\bibinfo  {journal} {Physical Review X}\ }\textbf {\bibinfo {volume} {10}},\
  \bibinfo {pages} {041002} (\bibinfo {year} {2020})}\BibitemShut {NoStop}%
\bibitem [{\citenamefont {Sakakibara}\ \emph {et~al.}(2020)\citenamefont
  {Sakakibara}, \citenamefont {Usui}, \citenamefont {Suzuki}, \citenamefont
  {Kotani}, \citenamefont {Aoki},\ and\ \citenamefont
  {Kuroki}}]{PhysRevLett.125.077003}%
  \BibitemOpen
  \bibfield  {author} {\bibinfo {author} {\bibfnamefont {H.}~\bibnamefont
  {Sakakibara}}, \bibinfo {author} {\bibfnamefont {H.}~\bibnamefont {Usui}},
  \bibinfo {author} {\bibfnamefont {K.}~\bibnamefont {Suzuki}}, \bibinfo
  {author} {\bibfnamefont {T.}~\bibnamefont {Kotani}}, \bibinfo {author}
  {\bibfnamefont {H.}~\bibnamefont {Aoki}},\ and\ \bibinfo {author}
  {\bibfnamefont {K.}~\bibnamefont {Kuroki}},\ }\bibfield  {title} {\bibinfo
  {title} {Model construction and a possibility of cupratelike pairing in a new
  ${d}^{9}$ nickelate superconductor
  $(\mathrm{Nd},\mathrm{Sr}){\mathrm{nio}}_{2}$},\ }\href
  {https://doi.org/10.1103/PhysRevLett.125.077003} {\bibfield  {journal}
  {\bibinfo  {journal} {Phys. Rev. Lett.}\ }\textbf {\bibinfo {volume} {125}},\
  \bibinfo {pages} {077003} (\bibinfo {year} {2020})}\BibitemShut {NoStop}%
\bibitem [{\citenamefont {Kita}\ \emph {et~al.}(2009)\citenamefont {Kita},
  \citenamefont {Ohashi},\ and\ \citenamefont
  {Suga}}]{kitaSpatialFluctuationsSpin2009}%
  \BibitemOpen
  \bibfield  {author} {\bibinfo {author} {\bibfnamefont {T.}~\bibnamefont
  {Kita}}, \bibinfo {author} {\bibfnamefont {T.}~\bibnamefont {Ohashi}},\ and\
  \bibinfo {author} {\bibfnamefont {S.-i.}\ \bibnamefont {Suga}},\ }\bibfield
  {title} {\bibinfo {title} {Spatial fluctuations of spin and orbital nature in
  the two-orbital {{Hubbard}} model},\ }\href
  {https://doi.org/10.1103/PhysRevB.79.245128} {\bibfield  {journal} {\bibinfo
  {journal} {Physical Review B}\ }\textbf {\bibinfo {volume} {79}},\ \bibinfo
  {pages} {245128} (\bibinfo {year} {2009})}\BibitemShut {NoStop}%
\bibitem [{\citenamefont {Nomura}\ \emph {et~al.}(2015)\citenamefont {Nomura},
  \citenamefont {Sakai},\ and\ \citenamefont
  {Arita}}]{nomuraNonlocalCorrelationsInduced2015}%
  \BibitemOpen
  \bibfield  {author} {\bibinfo {author} {\bibfnamefont {Y.}~\bibnamefont
  {Nomura}}, \bibinfo {author} {\bibfnamefont {S.}~\bibnamefont {Sakai}},\ and\
  \bibinfo {author} {\bibfnamefont {R.}~\bibnamefont {Arita}},\ }\bibfield
  {title} {\bibinfo {title} {Nonlocal correlations induced by {{Hund}}'s
  coupling: {{A}} cluster {{DMFT}} study},\ }\href
  {https://doi.org/10.1103/PhysRevB.91.235107} {\bibfield  {journal} {\bibinfo
  {journal} {Physical Review B}\ }\textbf {\bibinfo {volume} {91}},\ \bibinfo
  {pages} {235107} (\bibinfo {year} {2015})}\BibitemShut {NoStop}%
\bibitem [{\citenamefont {Georges}\ \emph {et~al.}(1996)\citenamefont
  {Georges}, \citenamefont {Kotliar}, \citenamefont {Krauth},\ and\
  \citenamefont {Rozenberg}}]{georgesDynamicalMeanfieldTheory1996}%
  \BibitemOpen
  \bibfield  {author} {\bibinfo {author} {\bibfnamefont {A.}~\bibnamefont
  {Georges}}, \bibinfo {author} {\bibfnamefont {G.}~\bibnamefont {Kotliar}},
  \bibinfo {author} {\bibfnamefont {W.}~\bibnamefont {Krauth}},\ and\ \bibinfo
  {author} {\bibfnamefont {M.~J.}\ \bibnamefont {Rozenberg}},\ }\bibfield
  {title} {\bibinfo {title} {Dynamical mean-field theory of strongly correlated
  fermion systems and the limit of infinite dimensions},\ }\href
  {https://doi.org/10.1103/RevModPhys.68.13} {\bibfield  {journal} {\bibinfo
  {journal} {Reviews of Modern Physics}\ }\textbf {\bibinfo {volume} {68}},\
  \bibinfo {pages} {13} (\bibinfo {year} {1996})}\BibitemShut {NoStop}%
\bibitem [{\citenamefont {Kotliar}\ \emph {et~al.}(2001)\citenamefont
  {Kotliar}, \citenamefont {Savrasov}, \citenamefont {P{\'a}lsson},\ and\
  \citenamefont {Biroli}}]{kotliarCellularDynamicalMean2001}%
  \BibitemOpen
  \bibfield  {author} {\bibinfo {author} {\bibfnamefont {G.}~\bibnamefont
  {Kotliar}}, \bibinfo {author} {\bibfnamefont {S.~Y.}\ \bibnamefont
  {Savrasov}}, \bibinfo {author} {\bibfnamefont {G.}~\bibnamefont
  {P{\'a}lsson}},\ and\ \bibinfo {author} {\bibfnamefont {G.}~\bibnamefont
  {Biroli}},\ }\bibfield  {title} {\bibinfo {title} {Cellular {{Dynamical Mean
  Field Approach}} to {{Strongly Correlated Systems}}},\ }\href
  {https://doi.org/10.1103/PhysRevLett.87.186401} {\bibfield  {journal}
  {\bibinfo  {journal} {Physical Review Letters}\ }\textbf {\bibinfo {volume}
  {87}},\ \bibinfo {pages} {186401} (\bibinfo {year} {2001})}\BibitemShut
  {NoStop}%
\bibitem [{\citenamefont {Maier}\ \emph {et~al.}(2005)\citenamefont {Maier},
  \citenamefont {Jarrell}, \citenamefont {Pruschke},\ and\ \citenamefont
  {Hettler}}]{maierQuantumClusterTheories2005}%
  \BibitemOpen
  \bibfield  {author} {\bibinfo {author} {\bibfnamefont {T.}~\bibnamefont
  {Maier}}, \bibinfo {author} {\bibfnamefont {M.}~\bibnamefont {Jarrell}},
  \bibinfo {author} {\bibfnamefont {T.}~\bibnamefont {Pruschke}},\ and\
  \bibinfo {author} {\bibfnamefont {M.~H.}\ \bibnamefont {Hettler}},\
  }\bibfield  {title} {\bibinfo {title} {Quantum cluster theories},\ }\href
  {https://doi.org/10.1103/RevModPhys.77.1027} {\bibfield  {journal} {\bibinfo
  {journal} {Reviews of Modern Physics}\ }\textbf {\bibinfo {volume} {77}},\
  \bibinfo {pages} {1027} (\bibinfo {year} {2005})}\BibitemShut {NoStop}%
\bibitem [{\citenamefont {Nomura}\ \emph {et~al.}(2014)\citenamefont {Nomura},
  \citenamefont {Sakai},\ and\ \citenamefont
  {Arita}}]{nomuraMultiorbitalClusterDynamical2014}%
  \BibitemOpen
  \bibfield  {author} {\bibinfo {author} {\bibfnamefont {Y.}~\bibnamefont
  {Nomura}}, \bibinfo {author} {\bibfnamefont {S.}~\bibnamefont {Sakai}},\ and\
  \bibinfo {author} {\bibfnamefont {R.}~\bibnamefont {Arita}},\ }\bibfield
  {title} {\bibinfo {title} {Multiorbital cluster dynamical mean-field theory
  with an improved continuous-time quantum {{Monte Carlo}} algorithm},\ }\href
  {https://doi.org/10.1103/PhysRevB.89.195146} {\bibfield  {journal} {\bibinfo
  {journal} {Physical Review B}\ }\textbf {\bibinfo {volume} {89}},\ \bibinfo
  {pages} {195146} (\bibinfo {year} {2014})}\BibitemShut {NoStop}%
\bibitem [{\citenamefont {Tubman}\ \emph {et~al.}(2020)\citenamefont {Tubman},
  \citenamefont {Freeman}, \citenamefont {Levine}, \citenamefont {Hait},
  \citenamefont {{Head-Gordon}},\ and\ \citenamefont
  {Whaley}}]{tubmanModernApproachesExact2020}%
  \BibitemOpen
  \bibfield  {author} {\bibinfo {author} {\bibfnamefont {N.~M.}\ \bibnamefont
  {Tubman}}, \bibinfo {author} {\bibfnamefont {C.~D.}\ \bibnamefont {Freeman}},
  \bibinfo {author} {\bibfnamefont {D.~S.}\ \bibnamefont {Levine}}, \bibinfo
  {author} {\bibfnamefont {D.}~\bibnamefont {Hait}}, \bibinfo {author}
  {\bibfnamefont {M.}~\bibnamefont {{Head-Gordon}}},\ and\ \bibinfo {author}
  {\bibfnamefont {K.~B.}\ \bibnamefont {Whaley}},\ }\bibfield  {title}
  {\bibinfo {title} {Modern {{Approaches}} to {{Exact Diagonalization}} and
  {{Selected Configuration Interaction}} with the {{Adaptive Sampling CI
  Method}}},\ }\href {https://doi.org/10.1021/acs.jctc.8b00536} {\bibfield
  {journal} {\bibinfo  {journal} {Journal of Chemical Theory and Computation}\
  }\textbf {\bibinfo {volume} {16}},\ \bibinfo {pages} {2139} (\bibinfo {year}
  {2020})}\BibitemShut {NoStop}%
\bibitem [{\citenamefont {{Williams-Young}}\ \emph {et~al.}(2023)\citenamefont
  {{Williams-Young}}, \citenamefont {Tubman}, \citenamefont {{Mejuto-Zaera}},\
  and\ \citenamefont {{de
  Jong}}}]{williams-youngParallelDistributedMemory2023}%
  \BibitemOpen
  \bibfield  {author} {\bibinfo {author} {\bibfnamefont {D.~B.}\ \bibnamefont
  {{Williams-Young}}}, \bibinfo {author} {\bibfnamefont {N.~M.}\ \bibnamefont
  {Tubman}}, \bibinfo {author} {\bibfnamefont {C.}~\bibnamefont
  {{Mejuto-Zaera}}},\ and\ \bibinfo {author} {\bibfnamefont {W.~A.}\
  \bibnamefont {{de Jong}}},\ }\bibfield  {title} {\bibinfo {title} {A
  parallel, distributed memory implementation of the adaptive sampling
  configuration interaction method},\ }\href
  {https://doi.org/10.1063/5.0148650} {\bibfield  {journal} {\bibinfo
  {journal} {The Journal of Chemical Physics}\ }\textbf {\bibinfo {volume}
  {158}},\ \bibinfo {pages} {214109} (\bibinfo {year} {2023})}\BibitemShut
  {NoStop}%
\bibitem [{\citenamefont {{Mejuto-Zaera}}\ \emph {et~al.}(2019)\citenamefont
  {{Mejuto-Zaera}}, \citenamefont {Tubman},\ and\ \citenamefont
  {Whaley}}]{mejuto-zaeraDynamicalMeanField2019}%
  \BibitemOpen
  \bibfield  {author} {\bibinfo {author} {\bibfnamefont {C.}~\bibnamefont
  {{Mejuto-Zaera}}}, \bibinfo {author} {\bibfnamefont {N.~M.}\ \bibnamefont
  {Tubman}},\ and\ \bibinfo {author} {\bibfnamefont {K.~B.}\ \bibnamefont
  {Whaley}},\ }\bibfield  {title} {\bibinfo {title} {Dynamical mean field
  theory simulations with the adaptive sampling configuration interaction
  method},\ }\href {https://doi.org/10.1103/PhysRevB.100.125165} {\bibfield
  {journal} {\bibinfo  {journal} {Physical Review B}\ }\textbf {\bibinfo
  {volume} {100}},\ \bibinfo {pages} {125165} (\bibinfo {year}
  {2019})}\BibitemShut {NoStop}%
\bibitem [{\citenamefont {Florez-Ablan}\ \emph {et~al.}(2025)\citenamefont
  {Florez-Ablan}, \citenamefont {Mejuto-Zaera},\ and\ \citenamefont
  {Capone}}]{bath_yo}%
  \BibitemOpen
  \bibfield  {author} {\bibinfo {author} {\bibfnamefont {D.}~\bibnamefont
  {Florez-Ablan}}, \bibinfo {author} {\bibfnamefont {C.}~\bibnamefont
  {Mejuto-Zaera}},\ and\ \bibinfo {author} {\bibfnamefont {M.}~\bibnamefont
  {Capone}},\ }\bibfield  {title} {\bibinfo {title} {Bath parameterization in
  multiband cluster dynamical mean-field theory},\ }\href
  {https://doi.org/10.1103/lnhv-kvy4} {\bibfield  {journal} {\bibinfo
  {journal} {Phys. Rev. B}\ }\textbf {\bibinfo {volume} {112}},\ \bibinfo
  {pages} {085153} (\bibinfo {year} {2025})}\BibitemShut {NoStop}%
\bibitem [{\citenamefont {{Mejuto-Zaera}}\ \emph {et~al.}(2020)\citenamefont
  {{Mejuto-Zaera}}, \citenamefont {{Zepeda-N{\'u}{\~n}ez}}, \citenamefont
  {Lindsey}, \citenamefont {Tubman}, \citenamefont {Whaley},\ and\
  \citenamefont {Lin}}]{mejuto-zaeraEfficientHybridizationFitting2020}%
  \BibitemOpen
  \bibfield  {author} {\bibinfo {author} {\bibfnamefont {C.}~\bibnamefont
  {{Mejuto-Zaera}}}, \bibinfo {author} {\bibfnamefont {L.}~\bibnamefont
  {{Zepeda-N{\'u}{\~n}ez}}}, \bibinfo {author} {\bibfnamefont {M.}~\bibnamefont
  {Lindsey}}, \bibinfo {author} {\bibfnamefont {N.}~\bibnamefont {Tubman}},
  \bibinfo {author} {\bibfnamefont {B.}~\bibnamefont {Whaley}},\ and\ \bibinfo
  {author} {\bibfnamefont {L.}~\bibnamefont {Lin}},\ }\bibfield  {title}
  {\bibinfo {title} {Efficient hybridization fitting for dynamical mean-field
  theory via semi-definite relaxation},\ }\href
  {https://doi.org/10.1103/PhysRevB.101.035143} {\bibfield  {journal} {\bibinfo
   {journal} {Physical Review B}\ }\textbf {\bibinfo {volume} {101}},\ \bibinfo
  {pages} {035143} (\bibinfo {year} {2020})}\BibitemShut {NoStop}%
\bibitem [{\citenamefont {Koch}\ \emph {et~al.}(2008)\citenamefont {Koch},
  \citenamefont {Sangiovanni},\ and\ \citenamefont
  {Gunnarsson}}]{kochSumRulesBath2008}%
  \BibitemOpen
  \bibfield  {author} {\bibinfo {author} {\bibfnamefont {E.}~\bibnamefont
  {Koch}}, \bibinfo {author} {\bibfnamefont {G.}~\bibnamefont {Sangiovanni}},\
  and\ \bibinfo {author} {\bibfnamefont {O.}~\bibnamefont {Gunnarsson}},\
  }\bibfield  {title} {\bibinfo {title} {Sum rules and bath parametrization for
  quantum cluster theories},\ }\href
  {https://doi.org/10.1103/PhysRevB.78.115102} {\bibfield  {journal} {\bibinfo
  {journal} {Physical Review B}\ }\textbf {\bibinfo {volume} {78}},\ \bibinfo
  {pages} {115102} (\bibinfo {year} {2008})}\BibitemShut {NoStop}%
\bibitem [{\citenamefont
  {S{\'e}n{\'e}chal}(2010)}]{senechalBathOptimizationCellular2010}%
  \BibitemOpen
  \bibfield  {author} {\bibinfo {author} {\bibfnamefont {D.}~\bibnamefont
  {S{\'e}n{\'e}chal}},\ }\bibfield  {title} {\bibinfo {title} {Bath
  optimization in the cellular dynamical mean-field theory},\ }\href
  {https://doi.org/10.1103/PhysRevB.81.235125} {\bibfield  {journal} {\bibinfo
  {journal} {Physical Review B}\ }\textbf {\bibinfo {volume} {81}},\ \bibinfo
  {pages} {235125} (\bibinfo {year} {2010})}\BibitemShut {NoStop}%
\bibitem [{\citenamefont {Capone}\ \emph {et~al.}(2004)\citenamefont {Capone},
  \citenamefont {Civelli}, \citenamefont {Kancharla}, \citenamefont
  {Castellani},\ and\ \citenamefont {Kotliar}}]{Capone1d}%
  \BibitemOpen
  \bibfield  {author} {\bibinfo {author} {\bibfnamefont {M.}~\bibnamefont
  {Capone}}, \bibinfo {author} {\bibfnamefont {M.}~\bibnamefont {Civelli}},
  \bibinfo {author} {\bibfnamefont {S.~S.}\ \bibnamefont {Kancharla}}, \bibinfo
  {author} {\bibfnamefont {C.}~\bibnamefont {Castellani}},\ and\ \bibinfo
  {author} {\bibfnamefont {G.}~\bibnamefont {Kotliar}},\ }\bibfield  {title}
  {\bibinfo {title} {Cluster-dynamical mean-field theory of the density-driven
  mott transition in the one-dimensional hubbard model},\ }\href
  {https://doi.org/10.1103/PhysRevB.69.195105} {\bibfield  {journal} {\bibinfo
  {journal} {Phys. Rev. B}\ }\textbf {\bibinfo {volume} {69}},\ \bibinfo
  {pages} {195105} (\bibinfo {year} {2004})}\BibitemShut {NoStop}%
\bibitem [{\citenamefont {Sakai}\ \emph {et~al.}(2012)\citenamefont {Sakai},
  \citenamefont {Sangiovanni}, \citenamefont {Civelli}, \citenamefont {Motome},
  \citenamefont {Held},\ and\ \citenamefont
  {Imada}}]{sakaiClustersizeDependenceCellular2012}%
  \BibitemOpen
  \bibfield  {author} {\bibinfo {author} {\bibfnamefont {S.}~\bibnamefont
  {Sakai}}, \bibinfo {author} {\bibfnamefont {G.}~\bibnamefont {Sangiovanni}},
  \bibinfo {author} {\bibfnamefont {M.}~\bibnamefont {Civelli}}, \bibinfo
  {author} {\bibfnamefont {Y.}~\bibnamefont {Motome}}, \bibinfo {author}
  {\bibfnamefont {K.}~\bibnamefont {Held}},\ and\ \bibinfo {author}
  {\bibfnamefont {M.}~\bibnamefont {Imada}},\ }\bibfield  {title} {\bibinfo
  {title} {Cluster-size dependence in cellular dynamical mean-field theory},\
  }\href {https://doi.org/10.1103/PhysRevB.85.035102} {\bibfield  {journal}
  {\bibinfo  {journal} {Physical Review B}\ }\textbf {\bibinfo {volume} {85}},\
  \bibinfo {pages} {035102} (\bibinfo {year} {2012})}\BibitemShut {NoStop}%
\bibitem [{\citenamefont {Bellomia}\ \emph {et~al.}(2024)\citenamefont
  {Bellomia}, \citenamefont {{Mejuto-Zaera}}, \citenamefont {Capone},\ and\
  \citenamefont {Amaricci}}]{bellomiaQuasilocalEntanglementMottHubbard2024}%
  \BibitemOpen
  \bibfield  {author} {\bibinfo {author} {\bibfnamefont {G.}~\bibnamefont
  {Bellomia}}, \bibinfo {author} {\bibfnamefont {C.}~\bibnamefont
  {{Mejuto-Zaera}}}, \bibinfo {author} {\bibfnamefont {M.}~\bibnamefont
  {Capone}},\ and\ \bibinfo {author} {\bibfnamefont {A.}~\bibnamefont
  {Amaricci}},\ }\bibfield  {title} {\bibinfo {title} {Quasilocal entanglement
  across the {{Mott-Hubbard}} transition},\ }\href
  {https://doi.org/10.1103/PhysRevB.109.115104} {\bibfield  {journal} {\bibinfo
   {journal} {Physical Review B}\ }\textbf {\bibinfo {volume} {109}},\ \bibinfo
  {pages} {115104} (\bibinfo {year} {2024})}\BibitemShut {NoStop}%
\bibitem [{\citenamefont {Kugel'}\ and\ \citenamefont
  {Khomskii}(1982)}]{kugelJahnTellerEffectMagnetism1982}%
  \BibitemOpen
  \bibfield  {author} {\bibinfo {author} {\bibfnamefont {K.~I.}\ \bibnamefont
  {Kugel'}}\ and\ \bibinfo {author} {\bibfnamefont {D.~I.}\ \bibnamefont
  {Khomskii}},\ }\bibfield  {title} {\bibinfo {title} {The {{Jahn--Teller}}
  effect and magnetism: Transition metal compounds},\ }\href
  {https://doi.org/10.1070/PU1982v025n04ABEH004537} {\bibfield  {journal}
  {\bibinfo  {journal} {Soviet Physics Uspekhi}\ }\textbf {\bibinfo {volume}
  {25}},\ \bibinfo {pages} {231} (\bibinfo {year} {1982})}\BibitemShut
  {NoStop}%
\bibitem [{\citenamefont {Park}\ \emph {et~al.}(2008)\citenamefont {Park},
  \citenamefont {Haule},\ and\ \citenamefont
  {Kotliar}}]{parkClusterDynamicalMean2008}%
  \BibitemOpen
  \bibfield  {author} {\bibinfo {author} {\bibfnamefont {H.}~\bibnamefont
  {Park}}, \bibinfo {author} {\bibfnamefont {K.}~\bibnamefont {Haule}},\ and\
  \bibinfo {author} {\bibfnamefont {G.}~\bibnamefont {Kotliar}},\ }\bibfield
  {title} {\bibinfo {title} {Cluster {{Dynamical Mean Field Theory}} of the
  {{Mott Transition}}},\ }\href
  {https://doi.org/10.1103/PhysRevLett.101.186403} {\bibfield  {journal}
  {\bibinfo  {journal} {Physical Review Letters}\ }\textbf {\bibinfo {volume}
  {101}},\ \bibinfo {pages} {186403} (\bibinfo {year} {2008})}\BibitemShut
  {NoStop}%
\bibitem [{\citenamefont {Balzer}\ \emph {et~al.}(2009)\citenamefont {Balzer},
  \citenamefont {Kyung}, \citenamefont {S{\'e}n{\'e}chal}, \citenamefont
  {Tremblay},\ and\ \citenamefont
  {Potthoff}}]{balzerFirstorderMottTransition2009}%
  \BibitemOpen
  \bibfield  {author} {\bibinfo {author} {\bibfnamefont {M.}~\bibnamefont
  {Balzer}}, \bibinfo {author} {\bibfnamefont {B.}~\bibnamefont {Kyung}},
  \bibinfo {author} {\bibfnamefont {D.}~\bibnamefont {S{\'e}n{\'e}chal}},
  \bibinfo {author} {\bibfnamefont {A.-M.~S.}\ \bibnamefont {Tremblay}},\ and\
  \bibinfo {author} {\bibfnamefont {M.}~\bibnamefont {Potthoff}},\ }\bibfield
  {title} {\bibinfo {title} {First-order {{Mott}} transition at zero
  temperature in two dimensions: {{Variational}} plaquette study},\ }\href
  {https://doi.org/10.1209/0295-5075/85/17002} {\bibfield  {journal} {\bibinfo
  {journal} {EPL (Europhysics Letters)}\ }\textbf {\bibinfo {volume} {85}},\
  \bibinfo {pages} {17002} (\bibinfo {year} {2009})}\BibitemShut {NoStop}%
\bibitem [{\citenamefont {Meixner}\ \emph {et~al.}(2024)\citenamefont
  {Meixner}, \citenamefont {Menke}, \citenamefont {Klett}, \citenamefont
  {Heinzelmann}, \citenamefont {Andergassen}, \citenamefont {Hansmann},\ and\
  \citenamefont {Sch{\"a}fer}}]{meixnerMottTransitionPseudogap2024}%
  \BibitemOpen
  \bibfield  {author} {\bibinfo {author} {\bibfnamefont {M.}~\bibnamefont
  {Meixner}}, \bibinfo {author} {\bibfnamefont {H.}~\bibnamefont {Menke}},
  \bibinfo {author} {\bibfnamefont {M.}~\bibnamefont {Klett}}, \bibinfo
  {author} {\bibfnamefont {S.}~\bibnamefont {Heinzelmann}}, \bibinfo {author}
  {\bibfnamefont {S.}~\bibnamefont {Andergassen}}, \bibinfo {author}
  {\bibfnamefont {P.}~\bibnamefont {Hansmann}},\ and\ \bibinfo {author}
  {\bibfnamefont {T.}~\bibnamefont {Sch{\"a}fer}},\ }\bibfield  {title}
  {\bibinfo {title} {Mott transition and pseudogap of the square-lattice
  {{Hubbard}} model: {{Results}} from center-focused cellular dynamical
  mean-field theory},\ }\href {https://doi.org/10.21468/SciPostPhys.16.2.059}
  {\bibfield  {journal} {\bibinfo  {journal} {SciPost Physics}\ }\textbf
  {\bibinfo {volume} {16}},\ \bibinfo {pages} {059} (\bibinfo {year}
  {2024})}\BibitemShut {NoStop}%
\bibitem [{\citenamefont
  {L{\"o}wdin}(1955)}]{lowdinQuantumTheoryManyParticle1955}%
  \BibitemOpen
  \bibfield  {author} {\bibinfo {author} {\bibfnamefont {P.-O.}\ \bibnamefont
  {L{\"o}wdin}},\ }\bibfield  {title} {\bibinfo {title} {Quantum theory of
  many-particle systems. {III}. extension of the {{Hartree-Fock}} scheme to
  include degenerate systems and correlation effects},\ }\href
  {https://doi.org/10.1103/PhysRev.97.1509} {\bibfield  {journal} {\bibinfo
  {journal} {Physical Review}\ }\textbf {\bibinfo {volume} {97}},\ \bibinfo
  {pages} {1509} (\bibinfo {year} {1955})}\BibitemShut {NoStop}%
\bibitem [{\citenamefont
  {L{\"o}wdin}(1964)}]{lowdinAngularMomentumWavefunctions1964}%
  \BibitemOpen
  \bibfield  {author} {\bibinfo {author} {\bibfnamefont {P.-O.}\ \bibnamefont
  {L{\"o}wdin}},\ }\bibfield  {title} {\bibinfo {title} {Angular momentum
  wavefunctions constructed by projector operators},\ }\href
  {https://doi.org/10.1103/RevModPhys.36.966} {\bibfield  {journal} {\bibinfo
  {journal} {Reviews of Modern Physics}\ }\textbf {\bibinfo {volume} {36}},\
  \bibinfo {pages} {966} (\bibinfo {year} {1964})}\BibitemShut {NoStop}%
\end{thebibliography}%
